\UseRawInputEncoding
\documentclass[reprint,superscriptaddress,amsmath,amssymb,aps,floatfix]{revtex4-2}

\usepackage{amsmath,amsfonts,amssymb,bm,cancel,multirow}
\usepackage{graphicx,float} 
\usepackage{enumitem}
\usepackage{color}
\usepackage{latexsym}
\usepackage{wasysym}
\usepackage{soul}
\usepackage{dcolumn}%
\usepackage{xcolor}
\usepackage{placeins}
\usepackage{old-tkz-graph}
\usetikzlibrary{arrows.meta}

\extrafloats{100}

\newcommand{\tagalign}[1]{\stepcounter{equation}\tag{\theequation}{#1}}

\newcommand{\nth}{$^{\text{th}}$\,}
\newcommand{\pp}[2]{\frac{\partial #1}{\partial #2}}
\newcommand{\dup}{\ensuremath{\mathrm d}}
\newcommand{\iup}{\ensuremath{\mathrm i}}

\newcommand{\hypergeometricF}{ {}_2F_1 }

\begin{document}

\title{Criticality in spreading processes without time-scale separation and the critical brain hypothesis}

\author{Daniel J. Korchinski}
\affiliation{Department of Physics and Astronomy, University of Calgary, Calgary, Alberta T2N 1N4, Canada}
\author{Javier G. Orlandi}
\affiliation{Department of Physics and Astronomy, University of Calgary, Calgary, Alberta T2N 1N4, Canada}
\author{Seung-Woo Son}
\affiliation{Department of Physics and Astronomy, University of Calgary, Calgary, Alberta T2N 1N4, Canada}
\affiliation{Department of Applied Physics, Hanyang University, Ansan, 15588, Republic of Korea}
\author{J\"orn Davidsen}
\affiliation{Department of Physics and Astronomy, University of Calgary, Calgary, Alberta T2N 1N4, Canada}
\affiliation{Hotchkiss Brain Institute, University of Calgary, Calgary, Alberta T2N 4N1, Canada}

\date{\today}

\begin{abstract}
Spreading processes on networks are ubiquitous in both human-made and natural systems. Understanding their behavior is of broad interest; from the control of epidemics to understanding brain dynamics. While in some cases there exists a clear separation of time scales between the propagation of a single spreading cascade and the initiation of the next --- such that spreading can be modelled as directed percolation or a branching process --- there are also processes for which this is not the case, such as zoonotic diseases or spiking cascades in neural networks. For a large class of relevant network topologies, we show here that in such a scenario the nature of the overall spreading fundamentally changes. This change manifests itself in a transition between different universality classes of \emph{critical} spreading, which determines the onset and the properties of an avalanche turning epidemic or neural activity turning epileptic, for example. We present analytical results in the mean-field limit giving the critical line along which scale-free spreading behaviour can be observed. The two limits of this critical line correspond to the universality classes of directed and undirected percolation, respectively. Outside these two limits, this duality manifests itself in the appearance of critical exponents from the universality classes of both directed and undirected percolation. We find that the transition between these exponents is governed by a competition between merging and propagation of activity, and identify an appropriate scaling relationship for the transition point. Finally, we show that commonly used measures, such as the branching ratio and dynamic susceptibility, fail to establish criticality in the absence of time-scale separation calling for a reanalysis of criticality in the brain.
\end{abstract}

\maketitle

\section{Introduction}
Diseases spreading on human contact networks~\cite{newman2002spread,kenah2007second,pastor2015epidemic}, worms and malware tunnelling through computer networks~\cite{wierman2004modeling}, rumours shared on social networks~\cite{nekovee2007theory}, power failures cascading on electrical networks~\cite{crucitti2004topological,wang2015network}, and neuronal avalanches in the brain~\cite{chialvo2010emergent,friedman2012universal,orlandi2013noise,tagliazucchi2012criticality,santo2018,kinouchi2006optimal,yaghoubi2018neuronal,dahmen2019} are all spreading processes unfolding on a network. Such processes typically exhibit a directed-percolation (i.e. branching process) phase transition as the probability of spreading passes a critical threshold. 

At this critical point, these systems exhibit shared scale-free statistics exhibiting characteristic power-laws. Neuronal spreading cascades or avalanches with similar critical exponents to that of directed percolation have been observed experimentally in neural systems that differ in size by many orders of magnitude, from \emph{in vitro} slices of a few hundred neurons, to whole-brain \emph{in vivo} calcium imaging and functional magnetic resonant imaging (fMRI)~\cite{beggs2003,chialvo2010emergent,friedman2012universal,tagliazucchi2012criticality,ponce2018}. These neuronal avalanches are at the core of the critical brain hypothesis, and link together self-organizing principles in brain dynamics and connectivity with optimal information processing~\cite{kinouchi2006optimal,dahmen2019}. Equivalent power-law distributions have been found to emerge naturally in the course of training artificial neural networks, suggesting they are a generic property of neural networks~\cite{del2017criticality}.

Although the mapping of neuronal avalanches to branching processes has proven quite successful, unavoidable discrepancies have appeared in recent years. There's an ongoing debate on whether these systems are really critical, quasi-critical, sub-critical, or a different definition altogether~\cite{Beggs2012,cocchi2017criticality}. This debate stems at least in part from challenges identifying a suitable order parameter and determining whether observed avalanches are truly power-laws following the predicted mean-field values.
Although an avalanche size distribution $p(S)\sim S^{-\tau}$ with $\tau \approx 1.5$ (consistent with mean-field) has been widely reported in the literature~\cite{beggs2003,Beggs2004,tagliazucchi2012criticality}, avalanches with exponents ranging from 1.2 to 2.5 have also appeared \cite{orlandi2013noise,moretti2013griffiths,yaghoubi2018neuronal,fontenele2019criticality,dalla2019modeling}. Whether the critical exponents even belong to directed percolation has also been challenged, with some proposing a oscillation-synchronization transition instead of a percolation transition \cite{poil2012critical,di2018landau,fontenele2019criticality,dalla2019modeling}. These observations are complicated by the experimental limitations of sub-sampling and coarse-graining \cite{pinheiro2019unified}. Further, experimental constraints make it challenging to distinguish sub-critical avalanches from those truncated by finite-size effects. 

Some of these issues however, might be due to the often overlooked fact that real systems rarely show proper time-scale separation; a classical branching process only allows for nodes to be excited when induced to do so by an antecedent node, and branching processes typically presuppose one ``root'' node. In other words, a branching process description assumes avalanches propagate and terminate on timescales much faster than the initiation of new avalanches.  Neural systems are not so simple however -- neurons can spontaneously activate due to ``minis'' or due to external sensory input. For smaller and intermediate systems, where multiple independent cascades are rare, the assumption of time-scale separation is not particularly limiting. If the rate of spontaneous activity is not too high and all activity is aggregated into a single avalanche, then $\tau$ changes from the expected 1.5 to 1.25~\cite{das2019critical}.  However, in large neural systems, there are no global quiet periods with which to delimit neuronal avalanches as in smaller systems. Some attempts at defining avalanches by activity exceeding a threshold exist, but it is not clear whether these thresholds identify a genuine critical point \cite{poil2012critical,di2018landau}.  This reflects the general challenge of defining criticality in strongly driven systems. However, some progress has been made.
To disentangle neural avalanches in whole-brain fMRI and study their statistics, Tagliazucchi et al. used physical proximity (i.e. nearest neighbour connections) to delimit avalanches, instead of binning all activity together~\cite{tagliazucchi2012criticality}. The same method was necessary to assess criticality in whole-brain zebra-fish data~\cite{ponce-alvarez_whole-brain_2018}. The approach of using network topology to identify avalanches makes sense, as information processing can only occur between connected elements of the network. To this end, neuronal avalanches have been generalized by ``causal webs'', which uses network structure to separate out independent avalanches~\cite{williams2017unveiling}. 

It has remained an open question whether a genuine critical point with scale-free activity can exist alongside spontaneous activity, particularly as other markers of a directed percolation transition (such as a unity branching ratio and the appearance of an active fraction) are all affected in different ways by spontaneous activity.  We address the issue here using a minimal spreading process with a spontaneous activation rate, making it a discrete-time $\varepsilon$-SIS (susceptible-infected-susceptible) model, and study it on a variety of complex networks.

Without spontaneous activations, there are no concurrent independent cascades of activity, and a directed percolation phase transition is present. We show that the introduction of spontaneous activations means that the macroscopic markers used to identify the directed percolation transition, (i.e. the appearance of an active fraction, or a branching ratio of one) no longer identify a phase transition. Nonetheless, by using the network structure to disentangle causally unrelated avalanches, we can define a phase transition even in the presence of spontaneous activations, with scaling-relations between exponents and finite-size scaling. We perform an extensive study of the critical properties at this phase transition on a variety of network structures and show that the presence of any spontaneous activation changes the underlying universality class from that of directed percolation to that of undirected percolation, while preserving some features of directed percolation. To explain these results, we derive an analytical mean-field theory for branching processes with spontaneous activation and show that the appearance of undirected percolation exponents is a direct result of the merging of initially independent avalanches. 

\section{Results}

\begin{figure}
	\begin{tikzpicture}[rotate=90,outline/.style={circle,draw,fill=white!40,minimum size=15}, spont/.style={circle,draw,fill=darkgray!40,minimum size=15}, activated/.style={circle,draw,fill=lightgray!40,minimum size=15}, EdgeStyle/.append style={->}]
	\pgfmathsetmacro{\xspc}{0.8}
	\pgfmathsetmacro{\tspc}{0.9}	
	\foreach \x in {0,...,4}
	{
		\pgfmathtruncatemacro{\labeltikz}{\x}
		\node [outline]  (\labeltikz) at (\xspc*\x,-\tspc*-1.5) {\small \labeltikz};
	}

	\foreach \x in {0,...,3}
	{
		\pgfmathtruncatemacro{\nextlabeltikz}{\x+1}
		\draw (\x) edge [->,bend left] (\nextlabeltikz);
		\draw (\nextlabeltikz) edge [->,bend left] (\x);		
	}

	\foreach \t in {0,...,6}{
		\foreach \x in {0,...,4}{
			\pgfmathtruncatemacro{\labeltikz}{10*\t + \x}
			\node [outline] (\labeltikz) at (\xspc*\x,-\tspc*\t) {};
		}
		\node (t\t) at (-\xspc,-\tspc*\t) {\small t=\t};
	}
	\node [spont,fill=black!60!blue] (s11) at (\xspc*1,-\tspc*1) {};
	\node [spont,fill=black!60!blue] (s24) at (\xspc*4,-\tspc*2) {};
	\node [spont,fill=black!60!red]   (s30) at (\xspc*0,-\tspc*3) {};
	\node [activated,fill=gray!40!blue] (a20) at (\xspc*0,-\tspc*2) {};
	\node [activated,fill=gray!40!blue] (a31) at (\xspc*1,-\tspc*3) {};
	\node [activated,fill=gray!40!blue] (a42) at (\xspc*2,-\tspc*4) {};
	\node [activated,fill=gray!40!blue] (a22) at (\xspc*2,-\tspc*2) {};
	\node [activated,fill=gray!40!blue] (a33) at (\xspc*3,-\tspc*3) {};
	\node [activated,fill=gray!40!red] (a41) at (\xspc*1,-\tspc*4) {};
	\node [activated,fill=gray!40!red] (a52) at (\xspc*2,-\tspc*5) {};
	
	\Edge (11)(20)
	\Edge (11)(22)
	\Edge (20)(31)
	\Edge (22)(31)
	\Edge (22)(33)
	\Edge (24)(33)
	\Edge (33)(42)
	\Edge (31)(42)
	
	\Edge (30)(41)
	\Edge (41)(52)
	\end{tikzpicture}
	
	\caption{Example dynamics of the branching process with spontaneous activations. Multiple spontaneous activations initiate on a simple linear bidirectional network (left). The dynamics here exhibit two independent avalanches, one with two roots (node 1 at time $t=1$ and node 4 at time $t=2$), and one with a single root (node 0 at time $t=3$). With spontaneous activations, spatially distinct events can overlap in time (e.g. $t=3$ and $t=4$) and initially distinct cascades of activity can overlap to form larger avalanches.   \label{fig:model}}
\end{figure}
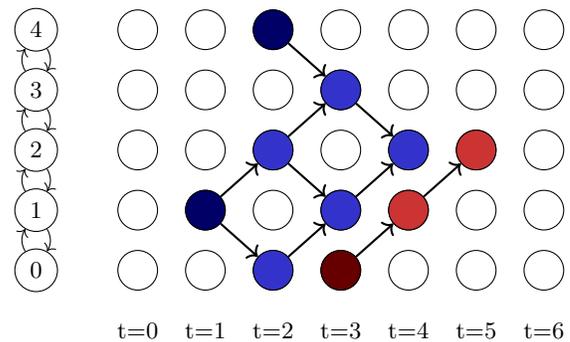

\subsection{Model} 
To study the effect of spontaneous activations, we consider here a discrete time $\varepsilon$-SIS process \cite{van2012epidemics} on directed networks equipped with spontaneous activations (see Fig.~\ref{fig:model}). At each time step, a given node can be activated by means of a spontaneous activation with probability $p$ or through an incoming link by an infected parent with probability $q$. More precisely, the probability that node $i$ is activated at time $t+1$ is given by:
\begin{equation}
    P(i,t+1) = 1 - (1-p)(1-q)^{m(i,t)},
    \label{eq:pactivation}
\end{equation}
where $m(i,t)$ counts how many parents of node $i$ were active at time $t$. Therefore spreading events occurs on a timescale of $1$, while spontaneous activations occur on a timescale of $p^{-1}$. This model be considered a type of Domany-Kinzel cellular automaton~\cite{hinrichsen2000nonequil}, or akin to some form of mixed-percolation~\cite{yanuka1990mixed}. In our model, nodes do not remain activated, but recover and so may be re-activated the following time step. In the Appendix~\ref{sec:appendix_SIR}, we also consider a variant of this model that includes the immunization of nodes, i.e., a susceptible-infected-recovered (SIR) process, with multiple initial spreaders.

Large systems with spontaneous activations will have concurrent and possibly unrelated avalanches. 
We employ the network structure itself to identify independent avalanches, using the ``causal-webs'' approach described in~\cite{williams2017unveiling}. To summarize the approach, we identify nodes with no active parents (i.e., no possible source of network-borne activation) as ``\emph{roots}'' of newly-initiated avalanches (see Fig.~\ref{fig:model}). Nodes with active parents inherit the avalanche labels of their parents. As avalanches overlap, they are merged together, so that nodes only ever have one label. This merging reflects that true causal information is often obscured in real systems, or that contributions from both streams of activity are necessary for activation.  This description of avalanches maps naturally to the clusters of traditional percolation.

The model has two limiting cases: i) For $p = 0$, this model is a pure branching process with branching parameter $q$, which belongs to the directed-percolation universality class. This case corresponds to neural activity where avalanches are infrequent and a single leading neuron can be positively identified in each avalanche. ii) $q = 0$ corresponds to the ordinary percolation model on a directed network with probability $p$. This corresponds to neural activity that is entirely driven by external sources or spontaneous activation and do not spread on the network, such as in the retina. Both limits exhibit continuous phase transitions, but fall into different universality classes, and are characterized by power laws exhibiting different critical exponents. Establishing that a phase transition exists for $p$ and $q$ that are simultaneously nonzero and the universality class of this transition are the principle efforts of this paper.

\subsection{Numerical results} 
\begin{figure}
    \includegraphics[width=\linewidth]{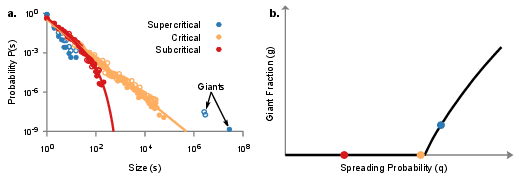}
	\caption{Approach to the phase transition in directed percolation. \textbf{a}. Avalanche distributions for with different spreading parameters $q$ with spontaneous activation rate $p \rightarrow 0$ on random $10$-regular networks.  Sub-critical avalanches are distributed with an exponentially-truncated power law $p(s) \sim s^{-3/2}\exp[-s/s_c]$. At criticality, avalanches are scale-free $s_c\rightarrow \infty$ such that $p(s)\sim s^{-3/2}$. The absolute size of the giants in the super-critical phase scales linearly with the network size (empty symbols $N=10^4$, filled $N=10^5$).  
	\textbf{b}. The giant fraction $g$ of nodes participating in the largest avalanche for the same simulations as in \textbf{a}. \label{fig:behaviour}}
\end{figure}

The most experimentally accessible indicator of criticality in systems with activity spreading is the size distribution of clusters, which shows a different critical exponent in directed and undirected percolation. For all $p$, at some critical $q_c(p)$ we observe a transition that defines a critical line. Below the critical point, with $q < q_c(p)$, avalanches are limited in size (see Fig.~\ref{fig:behaviour}{\bf a}), while above the critical point, a permanent giant component affecting a non-zero fraction of the network appears. 
At the critical point, the exponential cut-off that characterizes the sub-critical phase diverges and the avalanche distribution is described asymptotically by a power law. The appearance of these power laws is used to identify the critical line in our simulations. By studying the critical exponent characterizing these power laws, we can identify the universality class of the critical line.

\begin{figure}
	\includegraphics[width=\linewidth]{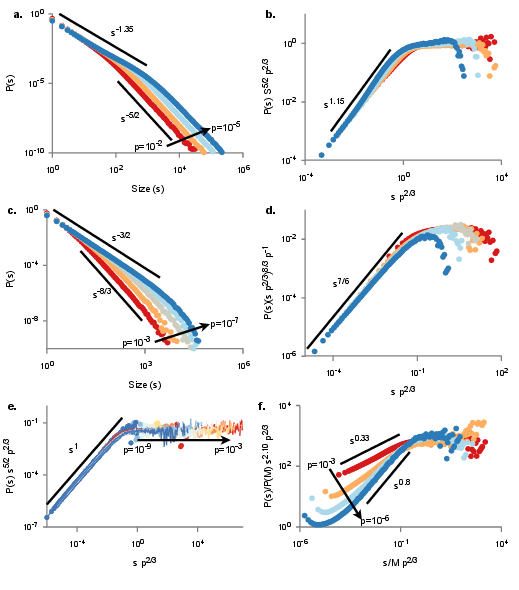}
	\caption{Critical avalanche statistics on critical line for various network topologies.  
	\textbf{a}. Small-world networks with $N=10^5$ and average degree $\langle k \rangle =10$. 
	\textbf{b}. As in {\textbf a}, but re-scaled to produce a curve collapse. 
	\textbf{c}. Uncorrelated power-law in- and out-degree distributions, $p(k) \sim k^{-3.5}$,  with $N = 10^7$ nodes. 	
	\textbf{d}. As in \textbf{c}, but rescaled to produce a curve-collapse. 
	\textbf{e}. Re-scaled avalanche distributions from 10-regular networks. Lines are simulations on infinite networks, while transparent circles are from finite simulations with $N=10^7$; both are at $q = q_c(p)$, the analytically-determined critical point. \textbf{f}. Re-scaled avalanche distributions from a hierarchical modular network, with $N = M\times 2^{15}$ nodes,  on a 15-layer hierarchy with base module size $M = 10^2$.   Solid lines are a guide for the eye. Unscaled panels \textbf{e}-\textbf{f} may be found in Figs.~\ref{fig:unscaled_avalanches}\textbf{a} and \ref{fig:unscaled_avalanches}\textbf{b}. The critical lines arising in each network topology are presented in Fig.~\ref{fig:phase_diagrams}. 	\label{fig:avdists}}
\end{figure}

Because critical exponents depend on the network topology and dimension, we study avalanche distributions with extensive simulations on a variety of relevant network architectures, including the small-world~\cite{song2014simple} and power-law networks relevant to  disease spreading (Figs.~\ref{fig:avdists}{\bf a}, \ref{fig:avdists}{\bf b}, \ref{fig:avdists}{\bf d}, and \ref{fig:avdists}{\bf e}), the hierarchical modular network~\cite{moretti2013griffiths} recently used as a brain connectome analogue (Fig.~\ref{fig:avdists}{\bf f}), and the analytically-tractable $k$-regular network (Fig.~\ref{fig:avdists}{\bf e}). Strikingly, for every level of spontaneous activity there is a transition between two power-law exponents in the critical avalanche distribution.

\begin{table*}
	\caption{ Calculated critical exponents for different network structures. In each column, we report both the theoretical value from percolation theory, and the value determined in this work, either analytically (by application of generating functions) or from numerical simulations (where they are reported with decimal values). Errors in $1/\overline{\nu}$ indicate the range over which an acceptable curve-collapse was obtained.
	A number of additional critical exponents were determined for the $k$-regular network, these are summarized in Table~\ref{tab:kreg_exponents}.  	\label{tab:exponent_table}}
	\begin{tabular}{c|c|c|c|c|c|c|c|c}
		Exponent & \multicolumn{2}{c|}{$\tau_{DP}$} & \multicolumn{2}{c|}{$\tau$} & \multicolumn{2}{c|}{$\beta$} & \multicolumn{2}{c}{${1}/{\overline{\nu}}$} \\ \hline 
		Quantity & \multicolumn{2}{c|}{$P(s) \sim s^{-\tau_{DP}}$ for $s<s_m$} & \multicolumn{2}{c|}{$P(s) \sim s^{-\tau}$ for $s>s_m$} & \multicolumn{2}{c|}{$g\sim (q-q_c)^\beta$} & \multicolumn{2}{c}{$q_c(N) - q_c \sim N^{-{1}/{\overline{\nu}}}$} \\ \hline 
		Small-World & - & $\approx 1.35$ & $5/2$~\cite{moore2000exact} & $\approx2.5$  &  1~\cite{moore2000exact} & $\approx 1.0$ & -&0.35(5) \\ \hline 
		Power-law   & 3/2~\cite{schwartz2002percolation} & $\approx 1.5$ & 8/3~\cite{cohen2002percolation}& $\approx 2.67$ & 2~\cite{cohen2002percolation} & $\approx 2.0$ & 1/5~\cite{cohen2002percolation} & 0.25(5) \\ \hline 
		Hierarchical Modular Network & Varies~\cite{munoz2010griffiths,moretti2013griffiths} & Varies & - & $\approx 2.1$ & - & $\approx 0.8$ &-&0.15(5) \\ \hline 
		$k$-Regular Network & 3/2~\cite{munoz1999avalanche} & 3/2 & 5/2~\cite{christensen2005complexity} & $\approx 2.5$ & 1~\cite{christensen2005complexity} & 1 & 1/3~\cite{cohen2002percolation} & 0.36(2) \\ 
	\end{tabular}
\end{table*}

We find that in all cases the first power-law exponent is consistent with the directed-percolation exponent for that system, while the latter exponent is indistinguishable from the pure percolation exponent (see Table~\ref{tab:exponent_table}). Yet, to the best of our knowledge, no directed-percolation avalanche exponent has been reported for directed small-world networks. To check the consistency of our findings for these networks, we can lower the density of long-range connections. Indeed, we find that the directed-percolation exponent for small-world networks tends toward the (1+1)-dimensional directed-percolation limit of $\approx 1.108$ expected of a circulant graph (see  Fig.~\ref{fig:sw_binder_1e-3} and associated text). For the power-law network, the corresponding degree exponent was chosen such that the undirected-percolation exponent would change (from the $5/2$ mean-field value to $8/3$ as predicted in~\cite{cohen2002percolation}) while leaving the directed-percolation exponent at $3/2$ (as predicted in~\cite{schwartz2002percolation}).

In hierarchical modular networks, we observe non-scale-free behaviour for avalanches below the base module size $M$ ($s<M$), a $p$ dependent power law for intermediate-size avalanches ($M<s<M p^{-2/3}$), and finally a single power law in the tail (cf. Fig.~\ref{fig:avdists}\textbf{f}). The varying exponent for intermediate-size avalanches is consistent with reports of a  Griffiths phase in modular networks with the SIS model~\cite{munoz2010griffiths,moretti2013griffiths,cota2018griffiths}  which belongs to the universality class of directed percolation. The largest avalanches are governed by an exponent of $\approx 2.1$, which matches with the undirected-percolation exponent for $q = 0$ (cf.  Fig.~\ref{fig:unscaled_avalanches}\textbf{b}). We hypothesize that this exponent is close to the pure 2-dimensional  percolation exponent, because the hierarchical modular network has a backbone that is very nearly one dimensional, and so the percolation process sees an effectively 2-dimensional lattice upon the introduction of time. All critical-avalanche distributions exhibit a universal curve collapse for various $p$ by re-scaling the distribution by $p^{-2/3}$  (Figs.~\ref{fig:avdists}{\bf b}, \ref{fig:avdists}\textbf{d}--\ref{fig:avdists}{\bf f}). This indicates that for all $p>0$, the critical point belongs to the same  universality class for that network topology.  
\begin{figure}
	\includegraphics[width=\linewidth]{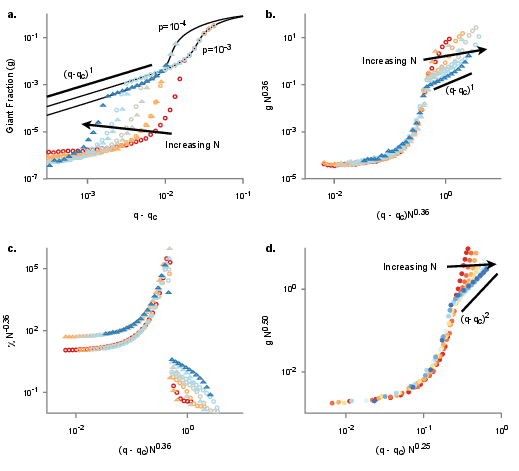}
	\caption{ Finite size scaling effects on giant component size and susceptibility. \textbf{a}. Giant components on 10-regular networks of varying sizes. Circles and triangles are  $p = 10^{-3}$ and $p = 10^{-4}$, respectively, while solid lines are the analytical calculations for infinite lattices. \textbf{b}. As in {\textbf a}, but re-scaled to produce a finite-size scaling curve collapse. \textbf{c}. Symbols are as in {\textbf a}, but studying the susceptibility $\chi = \langle s^2 \rangle_c$ for finite clusters, which shares the same finite-size scaling exponent. \textbf{d}. Curve collapse for power-law networks (degree distribution $p(k) \sim k^{-3.5}$) with $p = 10^{-4}$. Solid line is $g\sim (q - q_c)^\beta$ for $\beta = 2$, consistent with theoretical prediction.
	\label{fig:giants}}
\end{figure}
In the super-critical regime, a giant component appears, just as in directed and undirected percolation. The probability that a randomly-selected node is in the giant component is the giant component fraction $g$, which exhibits a power-law scaling, $g = \frac{G}{NT} \sim (q-q_c)^\beta$, where $G$ denotes the size of the largest cluster, $N$ the number of nodes in the system, and $T$ the simulation duration. However, $g$ exhibits a strong finite-size effect, with smaller systems having a larger effective critical point. Below the effective critical point, the largest cluster $G$ does not scale with the simulation duration and is not percolating. For this reason, we resort to finite-size scaling to reveal the critical behaviour of $g$ on finite networks (Fig.~\ref{fig:giants}). As $N\rightarrow \infty$, the effective critical point $q_c(N)$ tends towards the true critical point, with $(q_c(N) - q_c) \sim N^{-1/\bar{\nu}}$. The correctness of our finite-size scaling is confirmed by considering the finite cluster susceptibility  $\chi \equiv \langle s^2 \rangle_c$ (where the ${}_c$ denotes an average over all clusters), which should obey the same scaling collapse with $\chi N^{-\gamma / \bar{\nu}}$ (Fig.~\ref{fig:giants}{\bf c}). For mean field, ${1}/{\bar{\nu}} = {1}/{3}$ and for pure percolation on power-law networks (with degree distribution $p(k) \sim k^{-3.5}$), we expect that ${1}/{\bar{\nu}} = {1}/{5}$~\cite{cohen2002percolation}. Here, we find that the best scaling collapse occurs near these values, with $\frac{1}{\bar{\nu}} \approx 0.36$ for 10-regular networks and $\frac{1}{\bar{\nu}} \approx 0.25$ for power-law networks. 

Above the effective critical point, the giant component agrees with our analytical predictions for the infinite-size limit (Fig.~\ref{fig:giants}{\bf a}). As expected, the giant components emerge with $\beta = 1$ for the mean-field case of random $10$-regular networks (Figs.~\ref{fig:giants}{\bf a} and \ref{fig:giants}{\bf b}).   As can be seen in Fig.~\ref{fig:giants}{\bf d}, the giant component grows with $\beta = 2$ for the given power-law networks. Since $p(k)\sim k^{-3.5}$ and there are no correlations between the indegree and outdegree, it is known that $\beta = 2$ for undirected percolation~\cite{cohen2002percolation} and $\beta = 1$ for directed percolation~\cite{schwartz2002percolation}. This implies that the emerging giant component in our system is in the universality class of undirected percolation.

\subsection{Analytical results} 
In this section we establish analytically that the universality class of the phase transition is lifted from directed to undirected percolation by the addition of spontaneous activations. To understand the transition between directed- and undirected-percolation exponents, we consider the analytically tractable $k$-regular network. Using the generating function formalism, we can explicitly derive scaling exponents related to the avalanche size and emergence of the giant component, as well as derive a critical line. By studying the sizes of singly-rooted avalanches on this critical line, we can identify the size at which the merging of independent clusters of activity becomes the predominate mechanism for cluster growth, and thereby explain the scaling collapse effected by $p^{2/3}$ observed in Figs.~\ref{fig:avdists}{\bf c}--\ref{fig:avdists}{\bf f}. Additionally, the directed-percolation universality class exhibits two distinct diverging correlation lengths at the critical point. However, only one correlation length diverges in our model, reinforcing that this is an undirected-percolation transition. 

The avalanche distribution of our model is akin to the cluster-size distribution of percolation and directed percolation; this  distribution has been analytically determined on a variety of infinite random networks for both types of percolation by using probability generating functions (PGFs)~\cite{callaway2000network,newman2002spread,cohen2002percolation,schwartz2002percolation}. The technique's key assumption is that there are no loops and that all nodes are equivalent, in the sense that their network properties are independent of the properties of their neighbours. Although this is only an approximation, this tree-like approximation can still perform well in cases where loops are prevalent~\cite{melnik2011unreasonable}. This assumption lets one write down a self-consistent equation for the PGF in terms of the number of connected neighbours where the cluster that each neighbours connects to is distributed according to the original PGF. Our approach, detailed in Appendix~\ref{sec:appendix_gfunc}, follows that same spirit, except that a system of two self-consistently coupled PGFs are required to describe the total cluster distribution. One PGF corresponds to the sizes of clusters reached from a direct descendent, while the other describes the cluster size reached when two independent cascades merge. On a tree-like network, merging occurs when spontaneous activations meet and become a larger avalanche. Therefore, the directed-percolation-like behaviour is entirely contained within the first PGF, while the second PGF captures the effect of new spontaneous activations.

This pair of PGFs combines to define the PGF $H_0(x) = \sum_{s=1}^\infty P_n(s) x^s$ which corresponds to the avalanche-size distribution $P_n(s)$ obtained from sampling random active nodes (denoted with the sub-script $n$). Derivatives of $H_0(x)$ evaluated at $x=0$ directly yield $P_n(s)$. The average cluster size is just given by $\langle s\rangle_n = H_0'(1)$ and the susceptibility $\chi = \langle s^2 \rangle_c = {H_0'(1)}/{\int_0^1 H_0(x) \dup x}$, where the sub-script $c$ indicates sampling over clusters as opposed to active nodes. Since the giant component is the unique infinite-size avalanche, the probability a random active node belongs to it is just $1-H_0(1)$, and so the giant component is also determined by $H_0$. For sufficiently small $p$ and $q$, the giant component is zero and all clusters are finite, but as the critical line is approached the susceptibility $\chi$ diverges as $\chi \sim |q_c - q|^{-1}$ (for fixed $p$) or $\chi \sim |p_c - p|^{-1}$ (for fixed $q$) as derived in Appendix~\ref{sec:appendix_gfunc}. This divergence defines the critical line, which can be simply expressed as 
\begin{equation}
0 = k(1-\sigma)^2 - (k-1)\sigma \sigma_m\,, \label{eq:critical_line_simple}
\end{equation}
where $\sigma(p,q)$ is the reproduction number or branching ratio, and $\sigma_m(p,q)$ is the merging number, corresponding to the number of cascades of activity leading to a randomly-selected active node with at least one parent. Clearly then, the critical line has $\sigma < 1$ for all $\sigma_m > 0$, meaning that giants can occur even before an average reproduction number of $1$ is attained. 
\begin{figure}
	\includegraphics[width=\linewidth]{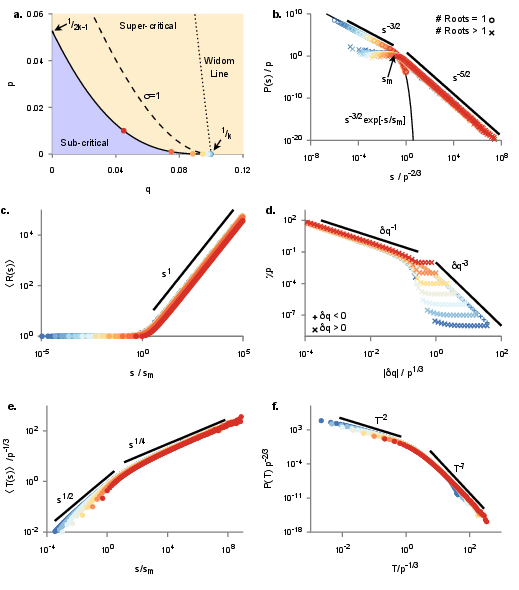}
	\caption{Power-law transitions are governed by merging. 
	\textbf{a}. Phase diagram for the $k$-regular network, with $k=10$. Points on the critical line correspond to the $(p,q_c(p))$in the other panels of this figure, with the $p$ ranging from $10^{-2}$ to $10^{-9}$. \textbf{b}. Re-scaled avalanche size distribution for various $p$ simulated on an infinite 10-regular network, partitioned into those avalanches with a single initiation site (empty circles) and those with multiple initiation sits (crosses). The theoretical distribution of mergeless avalanches is indicated with the solid line (cf. Eq.~(\ref{eq:mergeless_prob_asymptotic})).  \textbf{c}. Re-scaled average number of roots $R$ for avalanches of a given size for simulations on an infinite 10-regular network. 
	\textbf{d}. Re-scaled susceptibility near to the critical point, where $\delta q = q_c - q$, calculated by the generating function $H_0$. Sub-critical values $\delta q < 0$ are shown with empty circles and exhibit two power-laws, while super-critical values $\delta q > 0$ show only one. 
	\textbf{e}. Average avalanche duration for simulations of a given size collapse onto a single curve. 
	\textbf{f}. Avalanche duration distribution collapses onto a single curve with two power-laws. Unscaled data for panels \textbf{c}-\textbf{f} are found in Fig.~\ref{fig:unscaled_rollovers}.
		\label{fig:avroots}}	
\end{figure}

As shown in Fig.~\ref{fig:avroots}{\bf a}, $\sigma = 1$ only in the $p \rightarrow 0$ and $q\rightarrow \frac{1}{k}$ limit of directed percolation for a $k$-ary tree~\cite{christensen2005complexity}, where it agrees with the derived critical line. At the directed-percolation critical point, the active fraction of nodes $\Phi = \langle \Phi(t)\rangle$ exhibits a divergence in its dynamic susceptibility $\chi_0$, with $\chi_0 \equiv \pp{\Phi}{p}$ diverging as $\chi_0 \sim |\frac{1}{k} - q|^{-1}$. In the context of neural systems with mixed time-scales and a fixed level of spontaneous activation, the maximum of this dynamic susceptibility defines a ``Widom'' line (cf. Fig.~~\ref{fig:bp_susceptbility_widom} and associated text) and has been proposed as a quasi-critical line~\cite{williams2014quasicritical}. Although all three of these measures identify the directed-percolation critical point $p = 0$ and $q = \frac{1}{k}$, they disagree as soon as spontaneous activation is introduced ($p \neq 0$) and exhibit distinct scaling (cf. Fig.~\ref{fig:phasecurve_asymptote}). In the $p \ll 1$ limit, the  Widom line scales as $p \sim (\frac{1}{k} - q)$, the $\sigma = 1$ line scales as $p \sim (\frac{1}{k} - q)^2$ and the critical line scales as 
\begin{equation}
\left(\frac{1}{k} - q_c\right)^3 \approx \frac{(k-1)^2(2k-1)}{k^5}p_c \,. \label{eq:phasecurve_scaling}
\end{equation}
As for the $q =0$ endpoint to the critical line, $\chi$ diverges when $p = \frac{1}{(2k-1)}$ and $q = 0$, the pure percolation critical point for the Bethe lattice of coordination number $2k$. Hence, the critical line contains members belonging to two distinct universality classes.

To understand the appearance of the $p^{-2/3}$ scaling of the transition point shown in Fig.~\ref{fig:avdists}, we can consider the distribution of avalanches with only one root. These avalanches are described by a branching process, on a $k$-ary tree, with a branching probability $P_{d1} = (1-\Phi)^{k-1}(1- (1-p)(1-q))$ corresponding to the probability that a daughter branch activates with exactly one parent. The probability distribution for the size of the singly rooted avalanche is $P_\text{mergeless}(s) \sim s^{-3/2}\exp[-s / s_m]$, where $s_m  = -1 / \ln[ kP_{d1}(\overline{P_{d}}/(1-1/k))^{k-1}]$ (see Eq.~(\ref{eq:mergeless_sxi_exact})) denotes the characteristic scale above which avalanches merge and $\overline{P_{d}}$ is the probability a site does not activate, despite having an active parent.
Hence, we expect that the exponent $s^{-3/2}$ should be exponentially suppressed at $s_m$. In the limit of $p \to 0$ on the critical line (see Eq.~(\ref{eq:mergeless_sxi_goodprefactor})) $s_m$ scales as: 
\begin{equation}s_m \approx \frac{2(k-1)}{3k^3}\left(\frac{1}{k}-q\right)^{-2} \,. \label{eq:mergeless_scaling}\end{equation}
Combining Eqs. (\ref{eq:phasecurve_scaling}) and (\ref{eq:mergeless_scaling}) shows that the characteristic size before merging scales as $s_m \sim p^{-2/3}$ on the critical line. 

Simulations confirm that the smallest avalanches typically only have one root (Fig.~\ref{fig:avroots}{\bf b}), while the largest avalanches have a number of roots that scale with the avalanche size (Fig.~\ref{fig:avroots}{\bf c}). This means that there are two competing processes at play in these avalanches, both the propagation of the avalanche, which belongs in the directed-percolation universality class, and the merging of initially-independent events, which falls into the percolation universality class. This explains the appearance of the two power-laws and the associated curve collapse in Fig.~\ref{fig:avdists}. The first power-law is governed by the spreading of activity from a single initiation site, while the second power-law is governed by the merging of activity springing from multiple sites. This $-\frac{2}{3}$ scaling is a good approximation for random graphs that are close to mean-field. In Appendix~\ref{sec:appendix_smallworld}, we consider small-world networks with a low shortcut density. These networks are locally one-dimensional, and as a consequence exhibit a different scaling, $s_m \sim p^{-0.75}$ (cf.  Fig.~\ref{fig:sw_binder_1e-3}) due to the directed-percolation phase being (1+1)-dimensional instead of mean-field.

This transition between the directed- and undirected-exponents also manifests itself in the approach to the critical point. For instance, for $q < q_c$ the susceptibility can be approximated by
\begin{equation*}
\chi \approx \int_{1}^{s_m} s^2 s^{-\tau_{DP}} \dup s  + \Theta(s_\xi - s_m)\int_{s_m}^\infty s^2 s^{-\tau} F(s/s_\xi) \dup s \,,
\end{equation*}
where $\Theta$ is the Heaviside step function, and $s_\xi$ is the size cut-off of the pure-percolation tail, $s_\xi \sim |q_c - q|^{-1/\sigma}$ and $F$ is a universal scaling function. Then, using that $\frac{1}{k} - q_c \sim p^{1/3}$ from Eq.~(\ref{eq:phasecurve_scaling}), $\chi$ is (up to arbitrary multiplicative constants $C_1$, $C_2$), 
\begin{equation*}
\chi = C_1 p^{-1} (1 + \delta q / \sqrt[3]{p})^{3} + C_2\Theta(s_\xi - s_m)\delta q^{(3-\tau)/\sigma} \,.
\end{equation*}
This suggests that we see a transition between exponents when $\delta q \approx \sqrt[3]{p}$, precisely as observed in Fig.~\ref{fig:avroots}\textbf{d}. Now, since $\chi \sim \delta q^\gamma$ defines $\gamma$, we have arrived at the usual scaling relation $\gamma = \frac{3-\tau}{\sigma}$. This scaling relation holds for both the directed ($\gamma_{DP} = 3$, $\sigma_{DP} = \frac{1}{2}$, $\tau_{DP} = \frac{3}{2}$) and undirected ($\gamma = 1$, $\sigma = \frac{1}{2}$, and $\tau = \frac{5}{2}$) percolation regimes.

A transition from directed-percolation exponents also appears in the dynamical exponents relating the size of avalanches to their duration (cf. Fig.~\ref{fig:avroots}\textbf{e}), where the exponent transitions from $s\sim \langle T \rangle^{{\sigma \nu z} = \frac{1}{2}}$ to a power-law consistent with $s\sim T^\frac{1}{4}$. The onset of this transition again occurs with avalanches of size $s_m \sim p^{-\frac{2}{3}}$, which defines a characteristic time to merging, $T_m \sim \sqrt{s_m} \sim p^{-\frac{1}{3}}$. The scaling of this characteristic time captures an exponent transition in the distribution of the avalanche durations (cf. Fig.~\ref{fig:avdists}\textbf{f}, with $P(T)\sim T^\alpha$, with $\alpha_{DP} = 2$ and a new asymptotic $\alpha \approx 7.0$. Intriguingly, the directed-percolation scaling relation $\frac{\tau - 1}{\alpha - 1} = {\sigma \nu z}$ is satisfied even in the merging regime, assuming $\alpha = 7$, $\tau = 5/2$, and ${\sigma \nu z} = \frac{1}{4}$.

The existence of robust scaling relations and of curve collapses (Figs.~\ref{fig:avroots}\textbf{b}--\ref{fig:avroots}\textbf{f}) that appear universal indicate that the critical line (for $p > 0$) belongs to a single universality class. Since this includes the point $p = \frac{1}{2k-1}$ and $q = 0$, which we know is exactly undirected percolation, it suggests that the entire critical line (save for $p = 0$, $q = 1/k$) belongs to the universality class of undirected percolation. 

We can further strengthen the argument that the critical line is an undirected-percolation transition by studying the correlation lengths of the system. Undirected percolation exhibits a single isotropic diverging correlation length $\xi \sim  |\delta q|^{-\nu}$, while directed percolation exhibits two diverging correlation lengths  $\xi_\perp \sim |\delta q|^{-\nu_\perp}$ and $\xi_\parallel \sim |\delta q|^{-\nu_\parallel}$ corresponding to spatial and temporal correlation lengths respectively. We consider the correlation lengths corresponding to $\xi_\perp$ and $\xi_\parallel$ for our system, and will show that only one diverges on the critical line, precluding a directed-percolation transition. The two-point connectedness function, $\gamma(i,t_i,j,t_j)$ measures the probability that node $i$ at time $t_i$ and node $j$ at time $t_j$ belong to the same cluster over the ensemble average. If we denote the shortest path connecting nodes $i$ and $j$ as $d_{ij}$ then we expect that the average connectedness function should decay  with $d_{ij}$. This can be seen by studying the exponential decay of the average connectedness function, $g(d,t) = \langle \gamma(i,t_i,j,t_j) \rangle_{d_{ij} = d,\,t = t_j - t_i,\, \text{$i$ active}}$ which measures the decay of activity away from an active node. 

Typically, activity decays exponentially, with $g(2d,0) \sim \exp[-d / \xi_\perp]$ and $g(0,t) \sim \exp[-t/\xi_\parallel]$ defining the two correlation lengths $\xi_\perp$ and $\xi_\parallel$~\cite{hinrichsen2000nonequil}. In the loop-less (large $N$) approximation, $g(0,t) = \delta_{t0}$, as in the absence of loops activity can never return to the same site, meaning the correlation length $\xi_\parallel$ vanishes. Meanwhile, the perpendicular correlation length is given by $\xi_\perp = -1/\ln[\sigma^2(1+\sqrt{\beta})^2]$, which implies that the perpendicular correlation length diverges when $(1-\sigma)^2 = \sigma^2 \beta$, i.e. on the critical line where $\langle s \rangle_n$ diverges. The divergence of $\xi_\perp$ and the decay of $g(2d,0)$ is compared to its analytical form in  Fig.~\ref{fig:correlation_perp}. The non-divergence of $\xi_\parallel$ on the critical line suggests that the critical line is not a directed-percolation transition. One might object that the non-divergence of $\xi_\parallel$ is a problem with its construction. In the Appendix~\ref{sec:appendix_Correlation_length}, we consider also an isotropic correlation length that diverges on the critical line, and show that an alternative definition of $\xi_\parallel$ based on the typical number of generations of direct descendants to an active node only diverges on the $\sigma = 1$ line (cf. Fig.~\ref{fig:avroots}\textbf{a}). Additionally, the fact that $\xi_\perp$ diverges with $\nu_\perp = 1$, as in undirected percolation, instead of $\nu_\perp = \frac{1}{2}$ as expected in mean-field directed percolation reinforces that the critical line is an undirected-percolation transition.

In summary, the critical line is an undirected (as opposed to directed) percolation transition except at a singular point. This is supported by avalanche distribution exponents, exponents of the order parameter $g$, undirected-percolation scaling relations, and the divergence of a single correlation length. Many critical exponents of the directed percolation remain observable on small scales, such as in the beginning of the avalanche size distribution or in the susceptibility $\chi$. These exponents then shift to the undirected exponents when the merging of initially-independent avalanches becomes prevalent. Meanwhile, other measures of criticality that hold for directed percolation, such as the divergence of the dynamical susceptibility $\chi_0$ and a reproduction number of one, no longer capture critical behaviour. Instead, they predict phase-curves that agree only in the $p =0$ limit, and scale with different power-laws near the directed-percolation limit. Specifically, $\frac{1}{k} - q \sim p^{a}$, with $a = 1$ for the Widom line, $a= 2$ for the $\sigma=1$ line, and $a = 3$ for the critical line (c.f.  Fig.~\ref{fig:phasecurve_asymptote}). Thus, the directed-percolation transition is not robust with respect to the introduction of spontaneous activation -- any level of exogenous driving will introduce independent outbreaks, which on the largest scales will begin to merge. This is perhaps surprising, because the undirected-percolation limit $q = 0$ obeys detailed balance, while for the remainder of the critical line with $q > 0$ detailed balance is not respected.

\section{Discussion}
\subsection{General model observations}
We have described a two-parameter spreading process that includes spontaneous activations and exhibits a phase line along which the critical exponents and behaviour of both directed and undirected percolation appear. When there is no spontaneous activation, the model exhibits a directed-percolation transition, marked by a divergence in the dynamical susceptibility, a reproduction number of 1, power-law distributed outbreak sizes and the appearance of a giant component. However, the introduction of spontaneous activation means that the dynamical susceptibility no longer diverges, and that the reproduction number is shifted. Nonetheless, by considering the cluster-size distribution and statistics related to the cluster size, a critical line --- exhibiting universal curve collapses and finite-size scaling --- can be defined. This means that even in the presence of spontaneous activity, a genuine phase transition exists. The introduction of spontaneous activity destroys the transition to the absorbing state, and shifts the phase line into the universality class of undirected percolation. We showed numerically, on a variety of relevant network topologies, that in the largest clusters as merging becomes increasingly dominant it is the undirected-percolation exponents that dominate. Although all the networks we considered were nominally directed, we expect that our results survive on undirected networks, as our small-world networks were comprised predominately of bidirectional connections.

\subsection{Critical brain hypothesis}
Our results have repercussions for the critical brain hypothesis. Although the hypothesis itself is not new~\cite{bak1988self,Bienenstock1995}, it gained traction with the seminal work on neuronal avalanches~\cite{beggs2003,Beggs2004} and with the development of large-scale brain recording techniques. It is still a highly debated topic~\cite{Beggs2012,cocchi2017criticality}, and recent work has focused mostly on  (i) the appropriate definition of an order parameter and its tuning and (ii) whether neural activity distributions show critical power-law statistics.

Regarding (i) the main objective is to find a plausible mechanism by which the brain is able to tune its own activity to a critical point; ongoing research focuses on self-organized criticality~\cite{massobrio2015,Tetzlaff:2010811}, excitatory-inhibitory activity balance~\cite{Lombardi2012}, up and down states~\cite{millman2010,scarpetta2018}, adaptive mechanisms ~\cite{levina2007}, and learning~\cite{delpapa2017} amongst others. Many of these concepts are deeply related, and all of them might play a role. Regarding (ii), early work focused on whether the measured statistics followed a real power-law or just an approximation, and whether the activity was sub- or super-critical instead~\cite{Priesemann2013}. However, this is a challenging issue to solve experimentally due to the role of finite-size and sub-sampling effects~\cite{priesemann2014spike,levina2017,nonnenmacher2017}, and due to the real lack of separation of time scales as we report here. The initial reports on neuronal avalanches reported an exponent close to $\tau \approx 1.5$ for the size distributions, consistent with a mean-field branching process \cite{beggs2003,friedman2012universal},  but experiments on a variety of neural systems have reported a variety of exponents, usually in the range 1.2 - 2.5.  While this discrepancy might be partly explained by the technical challenges in probing the tails of power-law distributions, or attributed to to variation in network topology between studies and heterogeneous dynamical properties~\cite{yaghoubi2018neuronal}, there yet remain reasons to think spontaneous activation has a significant bearing on the critical brain hypothesis. For instance, critical \textit{in vivo} neuronal avalanches have only been reported for waking states, when animals presumably  are stimulated by sensory input~\cite{millman2010,Priesemann2013,scott_voltage_2014}. In light of our work, this might indicate that criticality is only attained with the help of external drive.

Our work addresses both (i) and (ii). For (i), we show that a susceptibility $\chi$ based on causal webs	accurately identifies the critical line, with the excellent finite-size scaling one expects from a true phase-transition. This offers a well-defined observable that is coherent in the present of spontaneous activity, unlike the branching ratio or a global measure like the dynamic susceptibility. As for (ii), we show that power-laws (with exponents that vary based on network structure) are present at the critical point for any level of spontaneous activation. While power-law statistics can also appear in non-critical systems with simpler dynamics as a recent critique showed~\cite{touboul2017}, the presence of scaling relations between the critical exponents~\cite{friedman2012universal} is only true in pure critical systems. We demonstrate such scaling relations in our model with coexisting power-law regimes, showing for the first time that neural networks with spontaneous activity can still be genuinely critical.  Additionally, our prediction of two power-laws may help to explain the variety of exponents fitted to power-laws in the literature.

Only recently other works have started to pay attention to the role of spontaneous activity for the critical brain hypothesis~\cite{williams2014quasicritical,williams2017unveiling,priesemann2018can,villegas2019thresholding,das2019critical,girardi2018,bolt2018}. Most of the previous definitions and requirements for criticality cannot be satisfied in the presence of spontaneous activity since time-scale separation is barely satisfied for any realistic activity rate. Several attempts have been made to recover the original definition of criticality and power-law statistics by accurately choosing and exploring the appropriate temporal bin size for the avalanche definition~\cite{yaghoubi2018neuronal}; but it is still not clear whether that approach really recovers the same underlying dynamics, and might only hold if the system is exactly at the critical point and for intermediate size systems. What is clear however, is that in the thermodynamic limit, with a fixed spontaneous activation rate, there will always be unrelated avalanches occurring, and so avalanches defined by global observables (such as fluctuations in $\Phi$ or those delimited by global quiescence) are not well-defined. As shown in the present work, in the absence of time-scale separation, it is essential to know the network structure to resolve the underlying dynamics. There is currently no way to recover the correct exponents without access to the network structure, but approaches that first try to infer the structure from the dynamics appear to be promising~\cite{rashid2017}.

Another key point that will have to be tackled in the future in relation to the critical brain hypothesis in the presence of spontaneous activity has to do with information transmission. Can it still be optimal in this regime? Spontaneous activity can indeed enhance information transmission from a sensory system~\cite{zierenberg2020}, or shift an inherently sub-critical system closer to the critical point (see Fig.~\ref{fig:avroots}). However, a read-out of this activity might require an error-correction code to be present.

\subsection{Aspects of disease spreading}
Much study of spreading on complex networks has been driven by a desire to study disease spreading in human contact networks, and we have borrowed heavily from that discipline in this paper. We would be remiss not to mention that a small number of studies have relaxed the patient-zero assumption of the typical spreading process, by including multiple initial spreaders~\cite{hu2014effects, miller2014epidemics, zheng2015non, hasegawa2016outbreaks, choi2017critical, hasegawa2018sudden, di2018multiple}. A limited few have  reflect a disease reservoir that can cause new outbreaks even as old ones spread \cite{van2012epidemics,cator2013susceptible,zhang2017contact}. To the best of our knowledge however, the question of the universality class of spreading as new spreaders are introduced has not yet been addressed in the literature. 

By endowing a spreading process with a spontaneous infection rate, we describe disease with an off-network reservoir. This might be an appropriate description for diseases like Zika virus, which can spread via human sexual networks, but also ``off-network'' via mosquito~\cite{rodriguez2016expanding}. Other zoonotic diseases lack an obvious patient zero, as they exhibit periodic reintroduction from animal reservoirs \cite{leroy2005fruit}. For example, two-thirds of new cases in the 2018 Democratic Republic of the Congo Ebola outbreak could not be linked to existing cases, potentially representing hidden network links or new outbreaks originating from contaminated bush meat~\cite{wolfe2005bushmeat,maxmen2018ebola}. In this case, the spontaneous infection rate could represent either infection from the environment, or a first order approximation for a failure in contact-tracing. 

Our model predicts that a non-zero spontaneous infection rate lowers the epidemic threshold. One consequence is that the local reproduction number can be less than one even in the epidemic phase. Further, for zoonotic diseases, the average outbreak size near the epidemic threshold will not follow the typical directed percolation scaling, but rather the isotropic percolation scaling exponents. Perhaps more importantly, close to the directed-percolation limit, a small change in the spontaneous infection rate can have a drastic impact on the average outbreak size. This suggests that a dual approach to epidemics, targeting both the transmission between individuals and the initial routes through which diseases enter the population, may represent a more efficient allocation of epidemiological intervention.  For example, one can imagine disease control as a constrained optimization problem attempting to minimize $\Phi$, with finite financial resources for disease interventions that affect $p$ and $q$. To compute Lagrange multipliers and find an optimal allocation requires calculating $\pp{\Phi}{p}$ and $\pp{\Phi}{q}$. We analytically do this for $k$-regular networks, but our approach can be directly adopted to networks more pertinent for disease spreading. By mapping real diseases onto our ``normal-form'', a concrete optimization problem can be constructed.

Finally, our SIS-based model describes a population that can be readily reinfected by a disease, i.e. with no acquired immunity. Although this is not representative of all diseases, the inclusion of immunization (as in the SIR model) does not impact our conclusions about the universality class of disease outbreaks with spontaneous infection. In the appendix (cf.  Fig.~\ref{fig:sw_static_model} and associated text), we consider a variant of our model that does not allow for the reinfection of nodes, and find outbreak distributions that exhibit a transition between directed- and undirected-percolation power-laws. This is a consequence of the fact that immunization only plays a role when a node would be reactivated i.e. after activity traverses a loop in the network. As we have shown with our tree-like (loop-free) analysis of the SIS case, these loops do not play a significant role in the mean-field limit. Immunization may play a stronger role in undirected networks or directed networks with many short-range loops. 

\subsection{Other applications}
The generality of our model makes it applicable to other systems where the timescales of spontaneous activation and propagation of activity are comparable, e.g., rumour spreading on social networks ~\cite{nekovee2007theory} or the distribution and propagation of computer viruses~\cite{wierman2004modeling}. It remains to be seen how our findings translate to self-organized systems and in particular to those that are known to exhibit a self-organized critical (SOC) regime under time-scale separation~\cite{jensen1998self,pruessner2012self,bak2013nature}. Previous studies have shown that a sufficiently high driving rate can induce a transition from avalanche dynamics to continuous flow in SOC systems~\cite{corral1999}. Within the context of zoonotic diseases, a multilayer network approach that incorporates human-animal interactions directly~\cite{Manilio2016NatPhys} or including cooperative diseases~\cite{Joaquin2014PRX} might open the door for novel dynamics in the absence of time-scale separation. It would also be interesting to see which metrics, such as $k$-shell decomposition~\cite{kitsak2010identification} or a local analysis~\cite{hu2018}, identify significant spreaders in our model and if network interventions, such as those connected to explosive percolation~\cite{nagler2011impact,grassberger2011explosive,d2015anomalous} and others~\cite{morone2017}, could be used to stymie or promote epidemics in the presence of spontaneous activation. These remain exciting challenges for the future.

\section{Conclusions}
Spreading processes on networks frequently appear in natural and human systems. The inclusion of spontaneous activity changes the phase transition in these systems from directed percolation to isotropic percolation, because previously independent streams of activity can merge together. These universality classes have differing critical exponents, meaning that diseases with spontaneous infections (e.g. zoonotic diseases) will show different growth profiles near the critical point. This also has several implications for the critical brain hypothesis. Global quantities -- such as the active fraction and its susceptibility, the branching ratio, or avalanches defined by global periods of quiescence -- do not capture critical behaviour when spontaneous activity is considered. As such, criticality in the brain should be re-assessed using measures that tolerate spontaneous activity. This requires that the use of network structure (e.g. tractography)  be paired with dynamics measurement (e.g. fMRI). Proximity to criticality should be assessed using an order parameter based on causal webs, such as susceptibility ($\chi$). Other measures of criticality, like the branching ratio, might lead to critical behaviour being interpreted as sub-critical. If the brain as a whole is critical, then the largest avalanches will have isotropic, rather than directed, percolation critical exponents as merging becomes the dominant growth mechanism. 

\begin{acknowledgments}
JD, DJK and JGO acknowledge helpful discussions with Rashid Williams-Garcia.
This project was financially supported by NSERC (JD, DJK, JGO), the Eyes High
Initiative of the University of Calgary (JGO, JD), Alberta Innovates (DJK) and the Korea-Canada Cooperative Development Program through the National Research Foundation of Korea (NRF) funded by the Ministry of Science and ICT, NRF-2018K1A3A1A74065535 (S-WS, JD) and Basic Science Research Program through NRF-2017R1D1A1B03032864 (S-WS). S-WS also acknowledges the support and hospitality of the Department of Physics \& Astronomy at the University of Calgary during his visit to Canada.
\end{acknowledgments}

\noindent
\appendix
\section{Methods}
\subsection*{Network generation}
For the finite networks, we generate finite directed $k$-regular networks via the configuration model, shuffling connections to avoid self-links and multi-links. To generate power-law networks we employed a variation of the configuration model described in~\cite{chen2013directed}, with a degree distribution  $p^{\text{in}/\text{out}}(k) = k^{-3.5}$ with a domain $k\in [5,...,1000]$, with rejection parameters $\kappa = 0.5$, $\delta = 0.05$. We generate the small-world networks using a directed network generalization of the Watts-Strogatz model~\cite{song2014simple}, using rewire probability $10^{-2}$ and with average degree 10 (a rewire probability of $10^{-3}$ is also considered in Appendix~\ref{sec:appendix_smallworld}). We generate the hierarchical modular networks described as ``HMN-2'' in~\cite{moretti2013griffiths} as a backbone for our modular networks. Within each base module with $n_c$ connections in the module backbone, we place $10^2+\alpha n_c$ nodes with $\alpha = 4$ being the four inter-modular connecting nodes. The first $M=10^2$ nodes we draw from the out-degree distribution $p(k) \sim e^{-(k-10)/(2\times0.5^2)}$ and connect to other uniformly drawn nodes in the same module. For the next $\alpha n_c$ nodes, we draw from the same out-degree distribution, but connect to the first $10^2$ nodes in the other modules, according to the module backbone wiring. 

\subsection*{Simulation of model on finite networks}
Networks are initiated with no active nodes. Each time-step, the number of nodes that will activate is drawn from the binomial distribution, with activation probability $p$. That number of nodes are randomly selected with uniform weighting, re-drawing duplicates. Nodes that activate spontaneously and had no active parents in the preceding time-step initiate a new cluster. Then, all nodes that had active parents in the preceding time step that were not already activated spontaneously are checked for activation. Each node with $m$ active parents in the previous time-step is activated with probability $1-(1-q)^m$. Nodes inherit the cluster label of their parents. If a node would inherit more than one cluster label, then those clusters are merged into a single cluster by relabelling all nodes belonging to the smaller cluster with the label of the larger cluster. Clusters that are found to have no active nodes in a given time-step are terminated, and their size, duration, and number of roots recorded.

\subsection*{Simulations on infinite $k$-regular networks} For the infinite networks, we begin at a randomly selected active node. We then check its immediate neighbours to see whether they are part of the same cluster. For those that are included, we then check their neighbours for inclusion. We can perform this process such that we need only count the number of unexplored neighbours, of which there are two types: (I) daughters that haven't been checked for inclusion and (II) parents that are known to be included, but whose neighbours haven't been checked.  If we're beginning from a root node, there are initially $k$ unchecked daughter branches (type I). If we're beginning from a randomly-active node, we begin with one type (I) neighbour and one type (II) neighbour. The algorithm proceeds to check each unevaluated connection (of type-I or type-II), possibly adding more as it goes, until none remain or the cluster exceeds a given size (typically $10^{10}$). Each type of connection is added as follows: 
\begin{itemize}
    \item \textbf{(Type I)}: We check each  type-I, by assuming it has $m_d$ other active parents (drawn from a binomial distribution $P(m_d) = \binom{k-1}{m_d}\Phi^{m_d}(1-\Phi)^{k-1-m_d}$ of $k-1$ other parents, activated with probability $\Phi$, the active fraction, given by Eq.~(\ref{eq:activefraction})), and include each type-I with probability $1-(1-p)(1-q)^{m_d}$. If it is included, then we add $k$ type-I connections from this daughter and $m_d$ type-II connections.
    \item \textbf{(Type II)}: Each type II is included with probability $1$. It adds $k-1$ additional type-I connections, and $m_p$ type-II parents, with $m_p$ drawn from the distribution in Eq.~(\ref{eq:pparents_probs}):
    \begin{equation}
        p(m_p) = \frac{\binom{k}{m_p} \Phi^{m_p} (1-\Phi)^{k-m_p}}{\Phi}(1 - (1-p)(1-q)^{m_p}). \label{eq:pparents_probs}
    \end{equation}
\end{itemize}
The probabilities of adding a daughter or parent are as derived in the appendix. For the purposes of measuring the two-point connectedness function, the above algorithm can be easily extended to also include the number of time-steps, by simply tracking how many times each active front has followed a daughter branch or a parent branch.

\begin{figure}[htbp]
	\includegraphics[width=\linewidth]{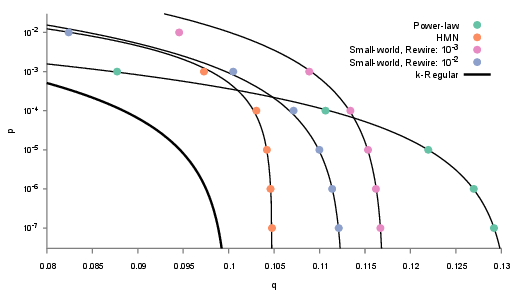}
	\caption{Numerically determined critical lines for various networks. Points correspond to the avalanche simulations plotted in Fig.~\ref{fig:avdists} in the main article, except for the small-world network with a re-wire of $10^{-3}$, which is only studied in the appendix. Solid lines are approximate fits of the form $p(q) = c(q_{c,DP} - q)^{a}$ for constant $c$, $a$, and $q_{c,DP}$, whose values are summarized in Table~\ref{tab:critical_fits}. $k$-regular phase curve is exact. \label{fig:phase_diagrams}}
\end{figure}

\subsection*{Critical point determination} Accurate determination of the critical point is necessary to effect accurate finite-size scaling. In the case of the $k$-regular network, the critical point can be determined analytically. However, for the power-law, small-world, and hierarchical modular networks, determination of the critical point can be done in two ways. The naive approach is to simply tune $q$ for fixed $N$ and $p$ until power-laws appear in the avalanche distribution. However, this is prone to finite-size effects: for fixed $p$ and $N$, the largest power-laws in the $P(s)$ distribution will appear at the pseudo-critical point, corresponding to a larger $q_c(N)$ value than the true $q_c(N \rightarrow \infty)$. Finite-size effects similarly cripple approaches based on just the appearance of the giant component or the diverging susceptibility.

\begin{table}
    \caption{Table of fit parameters for the critical lines of Fig.~\ref{fig:phase_diagrams}. Fits are power-law fits of the form $p = c(q_{c,DP}-q)^a$. Entries for the $k$-regular correspond to the low $p$ approximation given in Eq.~(\ref{eq:phasecurve_scaling}).}
    \begin{tabular}{c||c|c|c}
         & \multicolumn{3}{c}{Fit parameters} \\
        Network & c& a&$q_{C,DP}$  \\ \hline 
         Small-world, Rewire $=10^{-2}$ & 6.00 &2.97&0.112665 \\
         Small-world, Rewire $=10^{-3}$ & 7.99 &3.09&0.117104 \\ 
         Hierarchical Modular Network   & 2.78 &1.94&0.10482  \\ 
         Power-law network              & 2.29 &2.94&0.131118  \\ 
         $k$-Regular Network            & $ \frac{k^5}{(2k-1)(k-1)^2}$&3&$1/k$ \\ 
     \end{tabular}
    \label{tab:critical_fits}
\end{table}

Instead, we employ the method of de Souza et al.~\cite{de2011new} and consider the quantity $B = g \frac{\langle s^2 \rangle_c}{\langle s \rangle_c^2}$, which was shown to have no finite-size dependence at the critical point. Therefore, for fixed $p$, the critical point $q_c$ can be found as the intersection point of the $B(q)$ curves for different values of $N$ (see Fig.~\ref{fig:sw_binder_1e-3}\textbf{a}). This enables the numerical determination of the $p$, $q$ critical lines for the networks discussed in the main text (cf.  Fig.~\ref{fig:phase_diagrams} and Table~\ref{tab:critical_fits}).

\section{Generating Functions \label{sec:appendix_gfunc}}
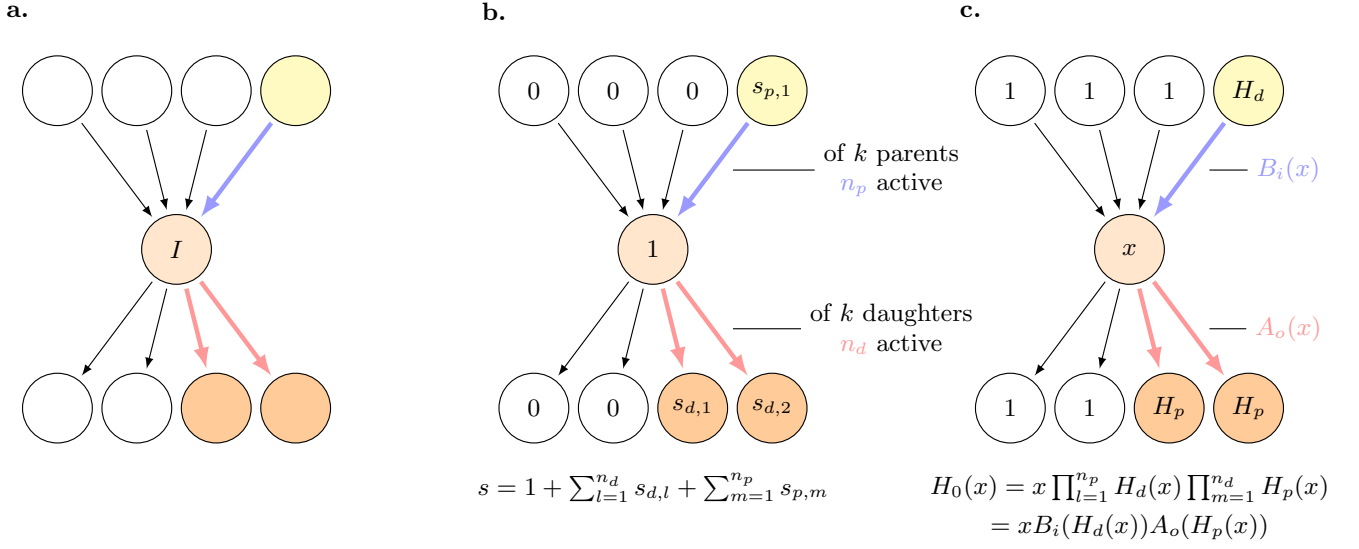
\begin{figure*}[htpb]
	\resizebox{\linewidth}{!}{
		\begin{tikzpicture}[outline/.style={circle,draw,fill=white!40,minimum size=25,outer sep=2},
		                    outlineLayer1/.style={circle,draw,fill=orange!20,minimum size=25,outer sep=2},
		                    outlineLayer0/.style={circle,draw,fill=yellow!30,minimum size=25,outer sep=2},
		                    outlineLayer2/.style={circle,draw,fill=orange!40,minimum size=25,outer sep=2}]		
		\node [outlineLayer1] (aCentreNode) at (0,0) {$I$};
		\node [outlineLayer2] (aDaughterNode) at (1.5,-2) {};
		\node [outlineLayer0] (aParentNode) at (1.5,2) {};
		\node [outline] (aD1) at (-1.5,-2) {};
		\node [outline] (aD2) at (-.5,-2) {};
		\node [outlineLayer2] (aD3) at (.5,-2) {};
		\node [outline] (aP1) at (-1.5,2) {};
		\node [outline] (aP2) at (-.5,2) {};
		\node [outline] (aP3) at (.5,2) {};
		\draw (aCentreNode) edge [-latex] (aD1);
		\draw (aCentreNode) edge [-latex] (aD2);
		\draw (aCentreNode) edge [-latex,ultra thick,red!40] (aD3);
		\draw (aCentreNode) edge [latex-] (aP1);
		\draw (aCentreNode) edge [latex-] (aP2);
		\draw (aCentreNode) edge [latex-] (aP3);
		\draw (aCentreNode) edge [-latex,ultra thick,red!40] (aDaughterNode);
		\draw (aCentreNode) edge [latex-,ultra thick,blue!40] (aParentNode);
		\node at (-2,3) {\textbf{a.}};
		
		\begin{scope}[shift = {(6,0)}]
		\node [outlineLayer1] (aCentreNode) at (0,0) {1};
		\node [outlineLayer2] (aDaughterNode) at (1.5,-2) {$s_{d,2}$};
		\node [outlineLayer0] (aParentNode) at (1.5,2) {$s_{p,1}$};
		\node [outline] (aD1) at (-1.5,-2) {0};
		\node [outline] (aD2) at (-.5,-2) {0};
		\node [outlineLayer2] (aD3) at (.5,-2) {$s_{d,1}$};
		\node [outline] (aP1) at (-1.5,2) {0};
		\node [outline] (aP2) at (-.5,2) {0};
		\node [outline] (aP3) at (.5,2) {0};
		\draw (aCentreNode) edge [-latex] (aD1);
		\draw (aCentreNode) edge [-latex] (aD2);
		\draw (aCentreNode) edge [-latex,ultra thick,red!40] (aD3);
		\draw (aCentreNode) edge [latex-] (aP1);
		\draw (aCentreNode) edge [latex-] (aP2);
		\draw (aCentreNode) edge [latex-] (aP3);
		\draw (aCentreNode) edge [-latex,ultra thick,red!40] (aDaughterNode);
		\draw (aCentreNode) edge [latex-,ultra thick,blue!40] (aParentNode);
		\node at (-2,3) {\textbf{b.}};
		\node (parentBox) [rectangle,minimum width = 0.5cm, minimum height = 1cm] at (.75,1) {};
		\node (parentLabel) [align=center] at (3.,1) {of $k$ parents \\ \textcolor{blue!40}{$n_p$} active};
		\draw (parentBox) edge [-] (parentLabel);
		
		\node (daughterBox) [rectangle,minimum width = 1cm, minimum height = 1cm] at (0.5,-1) {};
		\node (daughterLabel) [align=center] at (3.,-1) {of $k$ daughters \\ \textcolor{red!40}{$n_d$} active};
		\draw (daughterBox) edge [-] (daughterLabel);
		\node at (0,-3) {$s = 1 + \sum_{l = 1}^{n_d} s_{d,l} + \sum_{m=1}^{n_p} s_{p,m}$};
		\end{scope}
		
		\begin{scope}[shift = {(12,0)}]
		\node [outlineLayer1] (aCentreNode) at (0,0) {$x$};
		\node [outlineLayer2] (aDaughterNode) at (1.5,-2) {$H_p$};
		\node [outlineLayer0] (aParentNode) at (1.5,2) {$H_d$};
		\node [outline] (aD1) at (-1.5,-2) {1};
		\node [outline] (aD2) at (-.5,-2) {1};
		\node [outlineLayer2] (aD3) at (.5,-2) {$H_p$};
		\node [outline] (aP1) at (-1.5,2) {1};
		\node [outline] (aP2) at (-.5,2) {1};
		\node [outline] (aP3) at (.5,2) {1};
		\draw (aCentreNode) edge [-latex] (aD1);
		\draw (aCentreNode) edge [-latex] (aD2);
		\draw (aCentreNode) edge [-latex,ultra thick,red!40] (aD3);
		\draw (aCentreNode) edge [latex-] (aP1);
		\draw (aCentreNode) edge [latex-] (aP2);
		\draw (aCentreNode) edge [latex-] (aP3);
		\draw (aCentreNode) edge [-latex,ultra thick,red!40] (aDaughterNode);
		\draw (aCentreNode) edge [latex-,ultra thick,blue!40] (aParentNode);
		\node at (-2,3) {\textbf{c.}};	
		\node (parentBox) [rectangle,minimum width = 0.5cm, minimum height = 1cm] at (.75,1) {};
		\node (parentLabel) [align=center] at (2.,1) {\textcolor{blue!40}{$B_i(x)$}};
		\draw (parentBox) edge [-] (parentLabel);
		
		\node (daughterBox) [rectangle,minimum width = 1cm, minimum height = 1cm] at (0.5,-1) {};
		\node (daughterLabel) [align=center] at (2.,-1) {\textcolor{red!40}{$A_o(x)$}};
		\draw (daughterBox) edge [-] (daughterLabel);
		\node at (0,-3) {$H_0(x) = x \prod_{l=1}^{n_p} H_d(x) \prod_{m=1}^{n_d}H_p(x)$};
		\node at (0,-3.5) {$ = x B_i(H_d(x)) A_o(H_p(x))$};
		\end{scope}
		
		\end{tikzpicture}}
	\caption{A firing pattern example represented both as the sum of variables and the product of generating functions. \textbf{a.} An example activation pattern beginning from a randomly selected initial active node, I, on a $4$-regular network. Thick edges indicated connected active nodes. \textbf{b.} The number of parent and daughter edges contributing to the cluster are labelled by $n_p$ and $n_d$, random variables that could vary from $0$ to $k$. Nodes are labelled with their size contribution to the cluster -- nodes that do not activate contribute zero, the initially considered node contributes one activation, while active parents and daughter contribute a random variable. \textbf{c.} The same activity pattern labelled with the the probability generating functions corresponding to each random variable. \label{fig:tikz_h0_demo}}
\end{figure*}

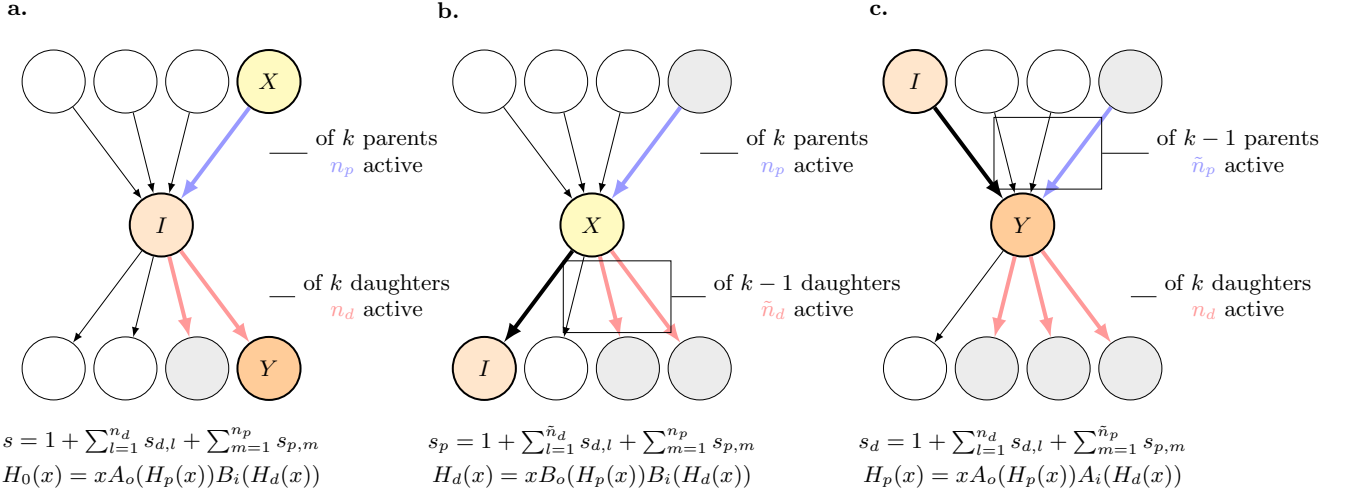
\begin{figure*}[htpb]
	\resizebox{\linewidth}{!}{
		\begin{tikzpicture}[
		    outline/.style={circle,draw,fill=white!40,minimum size=25},
		    outlineActive/.style={circle,draw,fill=gray!15,minimum size=25}
		    ]		
		\node [outline,thick,fill=orange!20] (aCentreNode) at (0,0) {$I$};
		\node [outline,thick,fill=orange!40] (aDaughterNode) at (1.5,-2) {$Y$};
		\node [outline,thick,fill=yellow!30] (aParentNode) at (1.5,2) {$X$};
		\node [outline] (aD1) at (-1.5,-2) {};
		\node [outline] (aD2) at (-.5,-2) {};
		\node [outlineActive] (aD3) at (.5,-2) {};
		\node [outline] (aP1) at (-1.5,2) {};
		\node [outline] (aP2) at (-.5,2) {};
		\node [outline] (aP3) at (.5,2) {};
		\draw (aCentreNode) edge [-latex] (aD1);
		\draw (aCentreNode) edge [-latex] (aD2);
		\draw (aCentreNode) edge [-latex,ultra thick,red!40] (aD3);
		\draw (aCentreNode) edge [latex-] (aP1);
		\draw (aCentreNode) edge [latex-] (aP2);
		\draw (aCentreNode) edge [latex-] (aP3);
		\draw (aCentreNode) edge [-latex,ultra thick,red!40] (aDaughterNode);
		\draw (aCentreNode) edge [latex-,ultra thick,blue!40] (aParentNode);
		
		\node (parentBox) [rectangle,minimum width = 3cm, minimum height = 1cm] at (0,1) {};
		\node (parentLabel) [align=center] at (3.,1) {of $k$ parents\\ \textcolor{blue!40}{$n_p$} active};
		\draw (parentBox) edge [-] (parentLabel);
		
		\node (daughterBox) [rectangle,minimum width = 3cm, minimum height = 1cm] at (0,-1) {};
		\node (daughterLabel) [align=center] at (3.,-1) {of $k$ daughters\\ \textcolor{red!40}{$n_d$} active};
		\draw (daughterBox) edge [-] (daughterLabel);
		
		\node at (0,-3) {$s = 1 + \sum_{l = 1}^{n_d} s_{d,l} + \sum_{m=1}^{n_p} s_{p,m}$};
		\node at (0,-3.5) {$H_0(x) = x  A_o(H_p(x)) B_i(H_d(x))$};
		\node at (-2,3) {\textbf{a.}};
		
		\begin{scope}[shift = {(6,0)}]
		\node [outline,thick,fill=yellow!30] (bCentreNode) at (0,0) {$X$};
		\node [outlineActive] (bDaughterNode) at (1.5,-2) {};
		\node [outlineActive] (bParentNode) at (1.5,2) {};
		\node [outline,thick,fill=orange!20] (aD1) at (-1.5,-2) {$I$};
		\node [outline] (aD2) at (-.5,-2) {};
		\node [outlineActive] (aD3) at (.5,-2) {};
		\node [outline] (aP1) at (-1.5,2) {};
		\node [outline] (aP2) at (-.5,2) {};
		\node [outline] (aP3) at (.5,2) {};
		\draw (bCentreNode) edge [-latex,ultra thick] (aD1);
		\draw (bCentreNode) edge [-latex] (aD2);
		\draw (bCentreNode) edge [-latex,ultra thick,red!40] (aD3);
		\draw (bCentreNode) edge [latex-] (aP1);
		\draw (bCentreNode) edge [latex-] (aP2);
		\draw (bCentreNode) edge [latex-] (aP3);
		\draw (bCentreNode) edge [-latex,ultra thick,red!40] (bDaughterNode);
		\draw (bCentreNode) edge [latex-,ultra thick,blue!40] (bParentNode);
		\node at (-2,3) {\textbf{b.}};
		
		\node (parentBox) [rectangle,minimum width = 3cm, minimum height = 1cm] at (0,1) {};
		\node (parentLabel) [align=center] at (3.,1) {of $k$ parents\\ \textcolor{blue!40}{$n_p$} active};
		\draw (parentBox) edge [-] (parentLabel);
		
		\node (daughterBox) [draw,rectangle,minimum width = 1.5cm, minimum height = 1cm] at (0.35,-1) {};
		\node (daughterLabel) [align=center] at (3.,-1) {of $k-1$ daughters\\ \textcolor{red!40}{$\tilde{n}_d$} active};
		\draw (daughterBox) edge [-] (daughterLabel);
		
		\node at (0,-3) {$s_p = 1 + \sum_{l = 1}^{\tilde{n}_d} s_{d,l} + \sum_{m=1}^{n_p} s_{p,m}$};
		\node at (0,-3.5) {$H_d(x) = x  B_o(H_p(x)) B_i(H_d(x))$};
		
		\end{scope}
		
		\begin{scope}[shift = {(12,0)}]
		\node [outline,thick,fill=orange!40] (bCentreNode) at (0,0) {$Y$};
		\node [outlineActive] (bDaughterNode) at (1.5,-2) {};
		\node [outlineActive] (bParentNode) at (1.5,2) {};
		\node [outline] (aD1) at (-1.5,-2) {};
		\node [outlineActive] (aD2) at (-.5,-2) {};
		\node [outlineActive] (aD3) at (.5,-2) {};
		\node [outline,thick,fill=orange!20] (aP1) at (-1.5,2) {$I$};
		\node [outline] (aP2) at (-.5,2) {};
		\node [outline] (aP3) at (.5,2) {};
		\draw (bCentreNode) edge [-latex] (aD1);
		\draw (bCentreNode) edge [-latex,ultra thick,red!40] (aD2);
		\draw (bCentreNode) edge [-latex,ultra thick,red!40] (aD3);
		\draw (bCentreNode) edge [latex-,ultra thick] (aP1);
		\draw (bCentreNode) edge [latex-] (aP2);
		\draw (bCentreNode) edge [latex-] (aP3);
		\draw (bCentreNode) edge [-latex,ultra thick,red!40] (bDaughterNode);
		\draw (bCentreNode) edge [latex-,ultra thick,blue!40] (bParentNode);
		\node at (-2,3) {\textbf{c.}};
		\node (parentBox)  [draw,rectangle,minimum width = 1.5cm, minimum height = 1cm] at (0.35,1) {};
		\node (parentLabel) [align=center] at (3.,1) {of $k-1$ parents\\ \textcolor{blue!40}{$\tilde{n}_p$} active};
		\draw (parentBox) edge [-] (parentLabel);
		
		\node (daughterBox) [rectangle,minimum width = 3cm, minimum height = 1cm] at (0,-1) {};
		\node (daughterLabel) [align=center] at (3.,-1) {of $k$ daughters\\ \textcolor{red!40}{$n_d$} active};
		\draw (daughterBox) edge [-] (daughterLabel);
		
		\node at (0,-3) {$s_d = 1 + \sum_{l = 1}^{n_d} s_{d,l} + \sum_{m=1}^{\tilde{n}_p} s_{p,m}$};
		\node at (0,-3.5) {$H_p(x) = x  A_o(H_p(x)) A_i(H_d(x))$};
		\end{scope}

		\end{tikzpicture}
	}
	\caption{An example of the three size generating functions. To illustrate the difference between $H_0$, $H_d$, and $H_p$ we consider the example firing pattern of Fig.~\ref{fig:tikz_h0_demo}, but centred on three different nodes, $X$, $I$, and $Y$. \textbf{a.} Corresponding to $H_0$, and  beginning from a randomly selected initial active node, $I$, on a $4$-regular network. Thick edges indicated connected active nodes. Two neighbouring active nodes, a parent and daughter ($X$ and $Y$ respectively) are highlighted as corresponding to the other two generating functions. \textbf{b.} An example activation pattern near $X$. Since one daughter connection leads to $I$, only $k-1$ are available for other connections. This restriction on daughters is why $H_d$ differs from $H_0$.  \textbf{c.} An example activation pattern near $Y$. Since one of the parents of $Y$ is $I$, only $k-1$ other parents need to be considered. In part due to the restrictions on  parents, $H_p$ differs from $H_0$. \label{fig:tikz_structure_demo}}
\end{figure*}
Before deriving the generating functions associated with the average cluster size, we begin with a brief review of probability generating functions (PGFs). For a discrete random variable $X$ drawn from the probability mass function $p(x)$, the probability generating function can be defined as 
\begin{equation*}
g_X(z) = E\left(z^X\right) = \sum_{x = 1}^\infty p(x) z^x \,.
\end{equation*}
$g_X$ generates the probability $p(x)$  in the sense that $g_X(0) = p(0)$, and the $n^{\text{th}}$ derivative yields: $\frac{1}{n!}g^{(n)}_X(0) = p(n)$.  The probability generating function can be used to obtain the moments of $X$, as $\langle X \rangle = g_X'(1)$, $\langle X(X-1) \rangle = g_X''(1)$, and so on.  The final property of probability generating functions we will use is perhaps its most useful: when a family of independent and identically distributed variables $\{ X_1, X_2, \ldots X_N \}$ generated by $g_X(z)$ are summed $Y = \sum^N X_i$, with $N$ also being a random variable generated by $g_N(z)$, then $g_Y(z) = E(z^Y) =  E(z^{NX})= \sum_{n=1}^\infty p(n=N)  \left( E\left(z^{X}\right) \right)^n = g_Y(g_X(z))$. Although this may seem esoteric, it means that the sum of a collection of some random number of random variables can be concisely expressed using generating functions. 

We will derive the PGFs corresponding to the cluster size distribution beginning from randomly selected active sites on a random network. If there are no loops in the network (the tree-like approximation), we can express the cluster size $s$ starting from a random active as 
\begin{equation}s = 1 + \sum_{l = 1}^{n_d} s_{d,l} + \sum_{m=1}^{n_p} s_{p,m}\,, \label{eq:h0_as_vars}\end{equation} where $n_d$ is the number of active daughters of the initial site, $n_p$ is the number of active parents of the initial site, $s_{d,l}$ is the size of the cluster reached from the $l$\nth active daughter, and $s_{p,m}$ is the size of the cluster reached from the $m$\nth  active parent. 
This means that the PGF for the total cluster size $s$ is
\begin{equation}
H_0(x) = x A_o(H_p(x))B_i(H_d(x)) \label{eq:h0_gfunc} \,,
\end{equation}
for the PGFs $A_o$ (generating $n_d$), $B_i$ (generating $n_p$), $H_p(x)$ (generating the $s_{d,l}$) and $H_d(x)$ (generating the $s_{p,m}$). The connection between the activation pattern and Eqs.~(\ref{eq:h0_as_vars},\ref{eq:h0_gfunc}) is illustrated in Fig.~\ref{fig:tikz_h0_demo}. 

An active daughter of $I$, here labelled $Y$, will have the number of parent branches that can be considered reduced by one, so,
\begin{equation*}
s_d = 1  + \sum_{l=1}^{\tilde{n}_p} s_{p,l}+ \sum_{m=1}^{{n}_d} s_{d,m}\,,
\end{equation*} where $\tilde{n}_p$ ranges from $0$ to $k-1$ and counts the parents other than $I$. $s_d$ is therefore generated by $H_p$ which obeys the following self-consistent equation:
\begin{equation}
H_p(x) =x A_o(H_p(x))A_i(H_d(x)) \label{eq:hp_selfcons} \,.
\end{equation}
Similarly, an active parent of $I$, here labelled $X$, will have one fewer daughter branch to consider, so its cluster size contribution is
\begin{equation*}
s_p = 1+ \sum_{l = 1}^{n_p} s_{p,l} + \sum_{m=1}^{\tilde{n}_d} s_{d,m} \,,
\end{equation*}
where $\tilde{n}_d$ ranges from $0$ to $k-1$, and counts the daughters other than $I$. $s_p$ is generated by $H_d$, which obeys the following self-consistent equation: 
\begin{equation}
H_d(x) = x B_o(H_p(x)) B_i(H_d(x)) \,. \label{eq:hd_selfcons} 
\end{equation} The relationship between the three size generating functions, $H_0$, $H_d$, and $H_p$ are illustrated in Fig.~\ref{fig:tikz_structure_demo}. 

To summarize, $H_p$ corresponds to the cluster size reached when arriving at a node from one of its \textbf{p}arent branches and $H_d$ corresponds to the cluster size reached when arriving at a node from one of its \textbf{d}aughter branches. The two pairs of generating functions ($A_i$, $A_o$) and ($B_i$, $B_o$) describe the number of active neighbours for $H_p$ and $H_d$ respectively. In terms of the nodes labelled in Fig.~\ref{fig:tikz_structure_demo}, the neighbour generating functions and their corresponding probability mass functions are: 
\begin{widetext}
\begin{align*}
A_i &\longleftrightarrow P(\tilde{n}_p\,\text{ active parents of $Y$ excluding $I$ }\,|\, \text{ $Y$ active \& $I$ active}) \tagalign{\label{eq:ai_pgf_pmf}} \\
A_o &\longleftrightarrow P(n_d \text{ active daughters of }Y \, |\, \text{ $I$ active}) \tagalign{\label{eq:ao_pgf_pmf}} \\
B_i &\longleftrightarrow P(n_p \text{ parents of }X\,|\,\text{$X$ active}) \tagalign{\label{eq:bi_pgf_pmf}} \\
B_o &\longleftrightarrow P(\tilde{n}_d \text{ active daughters excluding $I$ } \, |\, \text{ $X$ active})  \tagalign{\label{eq:bo_pgf_pmf}} 
\end{align*}
\end{widetext}

As the giant component appears, the average cluster size diverges. Therefore, identifying the conditions under which the average cluster size diverges is a natural way to identify the critical line. For $q\le q_c$, when $H_0(1) = H_p(1) = H_d(1) = 1$, the average cluster size is given by
\begin{equation}
\langle s\rangle_n = H_0'(1) = 1 + A_o'(1)H_p'(1) + B_i'(1)H_d'(1) \,, \label{eq:avg_size_generating}
\end{equation}
where the subscript $_{n}$ denotes an average conducted by sampling randomly selected nodes instead of averaging over clusters. Now, since $A_o'(1)$ and $B_i'(1)$ correspond to the mean number of daughters and parents of the initial randomly selected node, quantities that are necessarily bounded above by the mean (in/out)degrees, $A_o'(1)$ and $B_i'(1)$ cannot diverge. Therefore, $\langle s\rangle_n$ can only diverge if $H_p'(1)$ or $H_d'(1)$ do. Using Eqs.~(\ref{eq:hp_selfcons}) and (\ref{eq:hd_selfcons}), the following self-consistency relation for $H_p'(1)$ and $H_d'(1)$ (with $q\le q_c$) can be obtained.
\begin{equation}
\begin{bmatrix}
1-A_o'(1)&-A_i'(1) \\
-B_o'(1) & 1-B_i'(1) \\
\end{bmatrix} 
\begin{bmatrix}
H_p'(1)\\
H_d'(1)\\
\end{bmatrix} = \begin{bmatrix}
1 \\1
\end{bmatrix} \label{eq:HpHd_deriv}
\,.\end{equation}
Therefore $H_p'(1)$ and $H_d'(1)$ diverge when the determinant of the above matrix is zero, i.e., when $0 = (1-A_o'(1))(1-B_i'(1)) - B_o'(1)A_i'(1)$. This condition will yield the critical line, when supplied with the PGFs for $A$ and $B$. 

\subsection{Neighbour generating functions for k-regular networks}
So far, we've been quite generic in developing the generating function $H_0$. To proceed further, we must supply $A_{i/o}$ and $B_{i/o}$ for a given network. For simplicity, we focus on the $k$-regular network. This will allow us to develop expressions for $\Phi$, the active fraction, and $P_d$, the probability that the daughter of an active site activates in the next time step.  The first quantity we will need is the active fraction -- the proportion of nodes activated in each time step. A randomly selected (not necessarily active) node will have $m$ active parents with probability $\binom{k}{m} \Phi^m (1-\Phi)^{k-m}$,  as each parent is independent. With $m$ parents, the probability of activating is $1-(1-p)(1-q)^m$. Now since the probability of activation for a random node is also $\Phi$, we can write (using the notation $\overline{p} = 1-p$ to denote complementary probabilities) the self-consistent equation:
\begin{align*}
\Phi &= \sum_{m=0}^k  \binom{k}{m} \Phi^m \overline{\Phi}^{k-m} (1 - \overline{p}\cdot \overline{q}^m)\\
&=1-\overline{p}\cdot \overline{q \Phi}^k \, .\tagalign{\label{eq:activefraction}} 
\end{align*}
It will be useful, when performing asymptotic analysis in the limit that $p\rightarrow 0$, to have a closed-form approximation for $\Phi$. If we assume that $\Phi \ll 1$, we can truncate the expression $\overline{\Phi} =\overline{p}(1- k q \Phi + \frac{k(k-1)}{2} q^2\Phi^2 + \ldots)$ to first or second order in $\Phi$ and solve for $\Phi$, from which we obtain the first order approximation
\begin{equation}
    \Phi \approx \frac{p}{1 - kq} \label{eq:active_fraction_1storder}
\end{equation}
and the second order approximation (choosing the positive root, since $\Phi>0$)
\begin{equation}
\Phi \approx \frac{k \overline{p} q-1 + \sqrt{1 - k^2 \overline{p}^2 q^2- 2k\overline{p}q(1-p q)}}{(k-1)k \overline{p}q^2} \,. \label{eq:active_fraction_2ndorder}
\end{equation}
For $P_d$, we have one active parent, and $k-1$ parents that are independently active with probability $\Phi$. Hence,
\begin{align*}
P_d &= \sum_{m=0}^{k-1} \binom{k-1}{m}\Phi^m \overline{\Phi}^{k-1-m}\left(1-\overline{p}\cdot \overline{q}^{m+1}\right)\\
&= 1 - \overline{p}\cdot \overline{q} \cdot \overline{q \Phi}^{k-1}
\end{align*}
and simplifying using Eq.~(\ref{eq:activefraction})
\begin{equation}P_d = 1- \frac{\overline{q}\cdot \overline{\Phi}}{\overline{q \Phi}} \,.\label{eq:pdsimpleeq} \end{equation}
Note that $\sigma = k P_d$ defines the branching ratio.

Now that we have both $P_d$ and $\Phi$, we can derive $A_{i/o}$ and $B_{i/o}$. The simplest to derive are $A_o(x)$ and $B_o(x)$, because they describe the number of activated daughters, and the activation of each daughter is independent of the others. Considering a single daughter, whose activation can be described by a single random variable $m \in \{0,1\}$, with $m=1$ only if the single daughter activates. The PGF corresponding to $m$ is $C(x) = E(x^m) = (1-P_d)x^0 + P_d x^1 = \overline{P_d} + P_d x$. If $n$ is the number of activated daughters for a site with $l$ available daughters, then $n = \sum_{i=1}^l m_i$ for $m_l$ being independent and identically distributed (iid) Bernoulli variables generated by $C(x)$.  Then, taking $l=k$ for $A_o$, we have $A_o(x) = E(x^{n}) = E(x^{\sum_{i=1}^l m_i}) = \prod_{i=1}^{k}E(x^{m_i}) = C(x)^k$, so \begin{equation}A_o(x) = \left(\overline{P_d} + P_d x\right)^k\,. \end{equation} For $B_o$, we have one fewer daughter from which to choose, because we arrived at the node in question by means of one active daughter, so we take $l = k-1$ to find \begin{equation}B_o(x) = \left(\overline{P_d} + P_d x\right)^{k-1} \,.\end{equation}

Now, for $A_i$ and $B_i$, we cannot treat the parents' activation as independent. This is because we must condition on the knowledge that their daughter must activate and, in the case of $A_i$, also on the presence of other active parents.

Treating $B_i(x)$ first, we are considering an active site (labelled $X$) that we arrived at by means of an active daughter (in Fig.~\ref{fig:tikz_structure_demo}, $I$). Therefore, we have no knowledge about the number of active parents, save for the fact that they successfully activated the node in question. Considering the probability mass function in Eq.~(\ref{eq:bi_pgf_pmf}), Bayes' theorem allows us to write 
\begin{widetext}
\begin{equation*}P(n_p \text{ active parents of }X\,|\,\text{$X$ active}) = \frac{P(\text{$X$ active}\,|\,n_p\text{ active parents} ) P(n_p\text{ active parents} )}{P( \text{$X$ active})} \,. \end{equation*} 
\end{widetext}
However, $P(\text{X active}\,|\,n_p\text{ parents} ) = 1 - \overline{p} \, \overline{q}^{n_p}$ by definition of the model (Eq.~(\ref{eq:pactivation}) of the main text), while the probability of $n_p$ active parents, unconditioned on anything else is just given by $ P( n_p \text{ active parents} ) = \binom{k}{n_p} \Phi^{n_p} \overline{\Phi}^{k-n_p}$. Lastly, the probability that $X$ is active, conditioned on nothing else, is just the active fraction $\Phi$. Thus, 
\begin{align*} &P(n_p \text{ active parents of }X\,|\,\text{$X$ active}) \\ &\,= \frac{\left( 1 - \overline{p}\,\overline{q}^{n_p} \right) \left( \binom{k}{n_p} \Phi^{n_p} \overline{\Phi}^{k-n_p} \right)}{\Phi} \tagalign{\label{eq:si_p_np}}  \,,\end{align*}
which is exactly Eq.~(\ref{eq:pparents_probs}). The generating function corresponding to $B_i$ is therefore given by \begin{equation*}B_i(x) = \sum_{n_p=0}^k  P(n_p \text{ active parents of }X\,|\,\text{$X$ active}) x^{n_p}\end{equation*} and can therefore be expressed as  
\begin{equation}
B_i(x) = \frac{1}{\Phi}\left[ \left(\overline{\Phi}+ \Phi x \right)^{k} - \overline{p}\left(\overline{\Phi}+\Phi \overline{q}x\right)^k \right]
\end{equation}

For $A_i(x)$, we are considering a node, $Y$, that we arrived at from an active node (labelled $I$ in Fig.~\ref{fig:tikz_structure_demo}) that is one of $Y$'s parent branches. $A_i(x)$ is the generating function for the number of additional active parents of $Y$. Considering the probability mass function in Eq.~(\ref{eq:ai_pgf_pmf}), and applying Bayes' theorem, 
\begin{widetext}
\begin{align*} &P(\tilde{n}_p\,\text{ active parents of $Y$ excluding  $I$ }\,|\,\text{ $Y$ active}\,\&\,\text{ $I$ active} )  \\ &= {P(\text{ $Y$ active}\,|\,\text{ $I$ active}\,\&\,\text{$\tilde{n}_p$ other active parents of $Y$ })}   \times \frac{P( \text{$\tilde{n}_p$ of the $k-1$ parents other than $I$ active } )}{P(\text{$Y$ active}\,|\,\text{ $I$ active} )} \,.\end{align*}
Each of these probabilities are known. 
\begin{equation} P(\text{ $Y$ active}\,|\,\text{ $I$ active}\, \& \,\text{$\tilde{n}_p$ other parents of $Y$ active}) = 1 - \overline{p}\,\overline{q}^{\tilde{n}_p+1} \end{equation}
by definition of the model (Eq.~(\ref{eq:pactivation}) of main text), 
\begin{equation} P( \text{$\tilde{n}_p$ of $k-1$ parents other than $I$  active } ) = \binom{k-1}{\tilde{n}_p} \Phi^{\tilde{n}_p}\overline{\Phi}^{k-1-\tilde{n}_p}\,,\end{equation} and $P(\text{$Y$ active}\,|\,\text{ $I$ active} ) = P_d$. Hence, 
\begin{equation} P(\tilde{n}_p\,\text{ parents of $Y$ other than $I$ active}\,|\,\text{$Y$ active}\,\&\,\text{$I$ active}) = \frac{1}{P_d}\left(1-\overline{p}\,\overline{q}^{\tilde{n}_p+1}\right) \left(\binom{k-1}{\tilde{n}_p}\Phi^{\,\tilde{n}_p} \overline{\Phi}^{k-1-\tilde{n}_p}\right) \,.
\end{equation}
Now, the generating function $A_i$ is given by
\begin{equation*}A_i(x) = \sum_{\tilde{n}_p=0}^{k-1}x^{\tilde{n}_p} P(\tilde{n}_p\,\text{ active parents of $Y$ excluding $I$ }\,|\,\text{$Y$ active}\,\&\,\text{$I$ active})\,,\end{equation*} 
\end{widetext}
so after some algebra we have
\begin{equation*}
A_i(x) = \frac{1}{P_d}\left[\left(\overline{\Phi} + \Phi x\right)^{k-1} - \overline{p}\, \overline{q} \left(\overline{\Phi} + \Phi \overline{q}x\right)^{k-1} \right] \,.
\end{equation*}
This concludes the calculation of the four generating functions $A_{i/o}$ and $B_{i/o}$ for the $k$-regular network. These calculations can also be conducted for other random networks, although the calculation is more technically involved when the in-degree can vary or correlations exist between the in- and out-degrees. 

In summary, and in terms of $\Phi$ and $P_d$, the PGFs $A_{i/o}$ and $B_{i/o}$ for the $k$-regular network may be expressed as
\begin{equation}
A_o(x) = (\overline{P_d} + P_d x)^k \,, \label{eq:A_odefn}
\end{equation}
\begin{equation}
B_o(x) = (\overline{P_d} + P_d x)^{k-1} \,, \label{eq:B_odefn}
\end{equation}
\begin{equation}
A_i(x) = \frac{1}{P_d}\left[\left(\overline{\Phi} + \Phi x \right)^{k-1} - \overline{p}\,\overline{q}\left(\overline{\Phi} + \Phi \overline{q} x\right)^{k-1}\right] \,, \label{eq:A_idefn}
\end{equation}
and
\begin{equation}
B_i(X) = \frac{1}{\Phi}\left[\left(\overline{\Phi} + \Phi x \right)^{k-1} - \overline{p}\left(\overline{\Phi} + \Phi \overline{q} x\right)^k\right] \label{eq:B_idefn} \, .
\end{equation}

\subsection{Observables from the generating function}
Here, we summarize how to extract observables, such as the size fraction of the giant component $g$, susceptibility $\chi$, and cluster distribution $P_c(s)$ from the generating function $H_0(x)$. Practically speaking, we solve Eqs.~(\ref{eq:hp_selfcons}) and (\ref{eq:hd_selfcons}) self-consistently for $H_d(x)$ and $H_p(x)$ via a Newton-Raphson scheme for a given set of model parameters $p$, $q$, and $x$. With $H_d(x)$ and $H_p(x)$ in hand, we can insert these into Eq.~(\ref{eq:h0_gfunc}) and obtain $H_0(x)$. 

The first quantity we can obtain from $H_0(x)$ is the fraction of nodes involved in finite clusters, which is just $H_0(1) = \sum_s p(s)s = 1-P_\infty$. So the giant component fraction $g$, the fraction of all nodes at all times that are part of the infinite cluster, is just $g = \Phi (1-H_0(1))$. For the susceptibility, $\chi = \langle s^2 \rangle = \sum s^2 p_c(s)$, we must make the distinction between the cluster size distribution $p_c(s)$ (for numerical simulations, reported simply as $P(s)$) and the per-node cluster size distribution $P_n(s)$. The latter describes the cluster sizes observed by sampling random active nodes, and is directly calculated by the generating function approach, or accessed by simulating avalanches on the infinite lattice. Clearly, $P_n(s)=As P(s)$, for a normalization factor $A$. Since $\sum P(s) = 1$,  $A=\int_0^1 H_0(x) \dup x$. So, $\chi = \frac{1}{A}\sum s P_n(s)s  = \frac{\langle s\rangle_n}{A} = \frac{H_0'(1)}{\int_0^1 H_0(x) \dup x}$. Of course, we can directly access $P_n(s)$ by using $P_n(s) = \frac{1}{s!} \frac{\dup^s H_0(x)}{\dup x^s}\Bigr|_{x=0} $. As was pointed out in \cite{moore2000exact}, numerically evaluating this derivative for large $s$ is most easily accomplished via a contour integral 
\begin{equation}
\frac{\dup^s H_0(x)}{\dup x^s}\Bigr|_{x=0} = \frac{1}{2\pi \iup}\frac{ \oint H_0(z) dz }{z^{s+1}} \,,
\end{equation}
on the circle $z = e^{i\phi}$  for $\phi \in [0,2\pi]$. $z^d$ becomes highly oscillatory at large $d$, so convergence of this integral can be improved via standard numerical techniques for oscillatory integrals \cite{evans1999comparison}. The cluster probability distribution can then be accessed as $P(s) = \frac{1}{As}P_n(s)$. 

\subsection{Phase-diagram for the k-regular network}
We can study the divergence of $\chi \sim \langle s \rangle_n$ by solving Eq.~(\ref{eq:HpHd_deriv}), and inserting the solution into Eq.~(\ref{eq:avg_size_generating}) to obtain 
\begin{equation}\langle s \rangle_n = \frac{1 - \frac{\sigma}{k}
\left(\sigma - \sigma_m \right)}{(1-\sigma)^2 - \frac{k-1}{k}\sigma \sigma_m} \label{eq:avg_size_kreg} \end{equation}
where $\sigma_m = A_i'(1) = (k-1)P_{p1}  = (k-1)\frac{\Phi}{P_d} \left( 1 - \frac{(1-P_d)^2}{1-\Phi} \right)$ is the expected number of other active parents, to an active node with one already known parent. That is, $\sigma_m$ describes the rate of merging of initially independent clusters. Clearly, $\langle s \rangle_n$ diverges when $k(1-\sigma)^2 - (k-1)\sigma \sigma_m = 0$ (Eq.~(\ref{eq:critical_line_simple}) of the main text). This result could also have been arrived at by setting the determinant of Eq.~(\ref{eq:HpHd_deriv}) to zero. A reparameterization that will be convenient when considering the correlation length is to replace $\sigma_m$ with $\beta = \frac{(k-1)\sigma_m}{k \sigma}$, meaning that the critical line diverges when \begin{equation}(1-\sigma)^2 = \sigma^2 \beta\label{eq:critical_line_beta}\,.\end{equation} The set of $(p_c,\, q_c)$ that cause this divergence define a critical line (see Fig.~\ref{fig:avroots}a in the main text).

\begin{figure}[htbp]
\includegraphics[width=\linewidth]{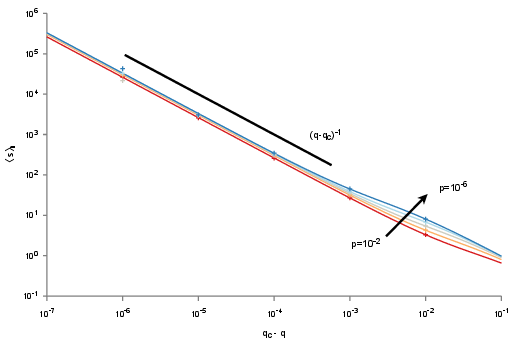}
\caption{Average cluster size for $k$-regular networks approaching critciality. Numerical simulations (points represent mean of $10^6$ realizations) on an infinite 10-regular graph yield good agreement with analytical predictions (lines) for $\langle s \rangle_n$ \label{fig:avg_s}}
\end{figure}
Solving $0 = (1-\sigma)^2  - \sigma^2 \beta$ for $q$, and assuming $\Phi \ll 1$ (as occurs in the $p \ll 1$ limit with $q < q_c$) yields $\frac{1}{k} - p_{1} = \frac{k-1}{k^2}\sqrt{2k-1}\sqrt{\Phi}$.  Inserting the first-order closed form approximation for $\Phi \ll 1$ (Eq.~(\ref{eq:active_fraction_1storder})) into the solution for $q$ yields the small $p$ expansion for the phase-curve (Eq.~(\ref{eq:phasecurve_scaling}) in the main text) \begin{equation} \left(\frac{1}{k}- q \right)^3  = \frac{(k-1)^2(2k-1)}{k^5} p \,.\end{equation}

The average cluster size $\langle s \rangle_n$ (and therefore susceptibility $\chi$) diverges for $(p,q)$ near to points on the critical line $(p_c,q_c)$ as $\langle s \rangle_n \sim |p_c - p|^{-\gamma}$ (for $q = q_c$) and $\langle s \rangle_n \sim |q_c - q|^{-\gamma}$ (for $p = p_c$) with $\gamma = 1$. This is a direct consequence of the fact that the numerator and denominator of Eq.~(\ref{eq:avg_size_kreg}) cannot both be simultaneously zero (except for the degenerate $q = 1$ case). Hence, the behaviour near the critical line will depend only on how the denominator $f(p,q) = (1-\sigma)^2 - \frac{k-1}{k}\sigma \sigma_m$ scales near its zero $p_c,q_c$. As $\pp{f}{p} \ne 0$ and $\pp{f}{q} \ne 0$ at $(p_c, q_c)$, the Taylor series approximation $f(p,q) \approx \pp{f}{p}( p -p_c )+ \pp{f}{q}  (q -q_c)$. Choosing $p = p_c$ or $q = q_c$ immediately yields the power-law scaling exponent $\gamma = 1$. This divergence can be visualized in Fig.~\ref{fig:avg_s}.

\subsection{The giant component}
The giant component fraction $g$ is given by $g = \Phi(1-H_0(1)) = \Phi \left( 1-H_d(1)\left[\overline{P_d} + P_d H_p(1) \right]\right)$. At the critical point, $H_p(1) = H_d(1) = 1$. So for $\delta = q - q_c \ll 1$, we have that $g \approx \left( \Phi P_d \pp{H_p}{q} + \Phi \pp{H_d}{q} \right) \delta $. Since both $H_d(1)$ and $H_p(1)$ are strictly decreasing functions of $q$, $g\sim (q - q_c)$ identifying the critical exponent $\beta = 1$.

\section{Mergeless Avalanches}
The mergeless clusters are exactly those clusters with one root. The number of configurations of singly rooted clusters of size $s$ is given by the Fuss-Catalan numbers  $C_s^{(k)} = \frac{1}{(k-1)s+1}\binom{ks}{s}$, which count the number of incomplete $k$-ary trees with $s$ vertices \cite{graham1989concrete}. Such a tree has perimeter (unoccupied branches) of length $t = (k-1)s+1$. Nodes are included in the tree with probability \begin{equation}P_{d1} = \overline{\Phi}^{k-1}(1- \overline{p} \,\overline{q})\,,\label{eq:pd1_simple}\end{equation} denoting the probability that a given daughter node is activated while having exactly one parent. The excluded nodes on the perimeter occur with probability $\overline{P_d}$, which is the probability of not activating, despite having an active parent. Hence, the probability of observing a mergeless cluster of size $s$ is given by $P(s) = C_s^{(k)}P_{d1}^{s-1}\overline{P_d}^t$. Applying Stirling's approximation we find \begin{equation} P(s) \sim s^{-3/2} \left( \frac{P_{d1} k}{((1-\frac{1}{k})/\overline{P_d})^{k-1}} \right)^s = s^{-3/2}e^{-s/s_m} \,, \label{eq:mergeless_prob_asymptotic}\end{equation} where the characteristic mergeless size is \[s_m = {-1} /{\log\left[ k P_{d1} \left( \frac{\overline{P_d}}{1-\frac{1}{k}}\right)^{k-1}\right]} \, .\] In the limit of $p \ll 1$ and $\frac{1}{k} - q \ll 1$, expressing $P_d$ (Eq.~(\ref{eq:pdsimpleeq})) and $P_{d1}$ (Eq.~(\ref{eq:pd1_simple})) in $p$, $q$, and $\Phi$, and again using the closed form approximation for $\Phi$ (Eq.~(\ref{eq:active_fraction_1storder})), we find to lowest order in $p$ and $(\frac{1}{k}-q)$:

\begin{table*}%
	\caption{Critical exponents for the $k$-regular network, beyond those reported in Table~\ref{tab:exponent_table} of the main text, where $\delta q = |q_c - q|$ and $ \delta q_{DP} = |q_{c,DP} - q|$, where $q_{c,DP}$ denotes the directed percolation critical point (i.e. at $p=0$). Exponents related to temporal dynamics (i.e. $\alpha$ and $1/\sigma \nu z$) reflect numerical observations from Fig.~\ref{fig:unscaled_rollovers}, while all other exponents are analytically determined. \label{tab:kreg_exponents}}
	\begin{tabular}{c|c|c|c}
		Exponent & Quantity & This work & Literature  \\ \hline 
		$\gamma_{DP}$ &$\chi = \langle s^2 \rangle_c \sim \delta q^{-\gamma_{DP}}$ for $\frac{\delta q}{p} \gg 1$& 3 & 3 \cite{munoz1999avalanche} (using $\gamma = \frac{3-\tau}{\sigma}$) \\ \hline 
		$\gamma$& $\chi \sim \delta q^{-\gamma}$ for $\frac{\delta q}{p} \ll 1$& 1 & 1 \cite{christensen2005complexity} \\ \hline 
		{$\alpha_{DP}$}&$P(T) \sim T^{-\alpha_{DP}}$ for $T<T_m$&$\approx 2$&  2 \cite{munoz1999avalanche} (using $\alpha = \delta+1$) \\ \hline 
		{$\alpha$}& {$P(T) \sim T^{-\alpha}$ for $T>T_m$}&$\approx 7$ & - \\ \hline 
		{$\frac{1}{\sigma \nu z_{DP}}$}&$T \sim s^{1/{\sigma \nu z_{DP}}}$ for $s<s_m$& $\approx 2$& 2 \cite{munoz1999avalanche} \\  \hline 
		{$\frac{1}{\sigma \nu z}$} &{$T \sim s^{\frac{1}{\sigma \nu z}}$ for $s>s_m$} & $\approx 4$&- \\ \hline  
		$\nu$ & $\xi \sim \delta q^{-\nu}$ & 1 &  1 \cite{christensen2005complexity}\\  \hline 
		$\frac{1}{\sigma_{DP}}$ & $s_m \sim \delta q_{DP}^{-1/\sigma_{DP}}$ & 2 & 2 \cite{munoz1999avalanche} \\ \hline 
		\multirow{2}{*}{$\frac{1}{\sigma}$} & $P(s) \sim s^{-\tau} G(s / s_\xi) $ for $s > s_m$ & \multirow{2}{*}{2} & \multirow{2}{*}{2 \cite{christensen2005complexity}} \\ 
		& and $G(x\gg 1) \rightarrow 0$ then $s_\xi \sim \delta q^{-\frac{1}{\sigma}}$ & & \\
	\end{tabular}
\end{table*}

\begin{equation}
    s_m^{-1} \approx \frac{(2k-1)(k-1)}{k^2} p\left(\frac{1}{k} -q\right)^{-1} + \frac{k^3}{2(k-1)} \left(\frac{1}{k} -q\right)^{2} \,, \label{eq:mergeless_sxi_exact}
\end{equation}
and if we apply Eq.~(\ref{eq:phasecurve_scaling}) to observe how the cut-off scales on the critical line, we find:
\begin{equation}
s_m \approx \frac{2(k-1)}{3 k^3} \left(\frac{1}{k} - q\right)^{-2} \,, \label{eq:mergeless_sxi_goodprefactor}
\end{equation}
which was Eq.~(\ref{eq:mergeless_scaling}) from the main text. Since $q_{c,DP}=\frac{1}{k}$ and the relation $s_m \sim \left(q_{c,DP} - q\right)^{-1/\sigma^{DP}}$ defines the directed percolation exponent $\sigma^{DP}$, we have also recovered the usual directed percolation exponent $\sigma^{DP} = \frac{1}{2}$ (cf. Table~\ref{tab:kreg_exponents}).

\section{Widom line}
\begin{figure}[htbp]
	\includegraphics[width=\linewidth]{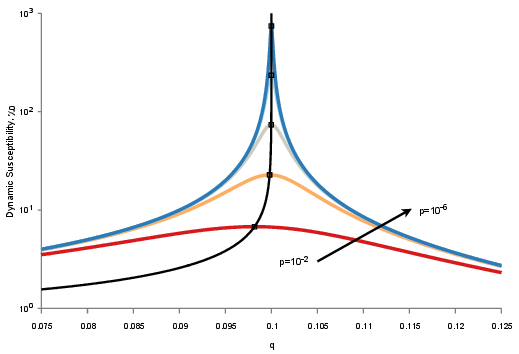}
	\caption{The dynamic susceptibility for various $p$ and $q$ as calculated for an infinite $10$-regular graph. For each $p$ there is a corresponding $q$ that maximizes the susceptibility. These maxima are labelled by the squares, and fall on the Widom-line.   \label{fig:bp_susceptbility_widom}}
\end{figure}
In equilibrium critical points, divergence in the correlation length is associated with a divergence in the susceptibility of the order parameter to an infinitesimal application of an external field. In directed percolation, the order parameter is $\Phi$. The susceptibility measures activity of the system in response to an external stimuli. We can imagine that the external stimuli is an infinitesimal increase in the average spontaneous activity of the system, and hence we can define the dynamic susceptibility as $\chi_0 \equiv \pp{\Phi}{p}$. So, using Eq.~(\ref{eq:activefraction}) we find:
\begin{align*}
\chi_0 &= \overline{\Phi q}^k + q k \overline{p}\,\overline{\Phi q}^{k-1} \chi_0 \,, \\
\implies \chi_0 &= \frac{\overline{p}\,\overline{\Phi q}^{k+1}}{\overline{p}(\overline{q\Phi}-qk \overline{p}\,\overline{q\Phi}^k)} \,, 
\end{align*}
and simplifying with Eq.~(\ref{eq:activefraction}) we obtain
\begin{equation}
\chi_0 = \frac{\overline{\Phi}\,\overline{q\Phi}}{\overline{p}(\overline{q\Phi} - q k \overline{\Phi})}  \,.
\end{equation}
In the limit $p \rightarrow 0$ with $\Phi \rightarrow 0$ and $q \rightarrow 1/k$, $\chi_0$ is, asymptotically, $\chi_0 \sim \frac{1}{k}\left(\frac{1}{k} - q\right)^{-1}$, and therefore diverges at the directed percolation critical point $q=\frac{1}{k}$ and $p=0$. This susceptibility has been studied in the context of neural systems, where the mixing of initiation and spreading time-scales means $\chi_0$ no longer diverges (cf. Fig.~\ref{fig:bp_susceptbility_widom}), but instead is maximized on a quasi-critical ``Widom'' line, where the fluctuations $\textrm{Var}(\Phi(t))$ are also maximized \cite{williams2014quasicritical}.

\section{Phase curve scaling}
\begin{figure}[htbp]
	\includegraphics[width=\linewidth]{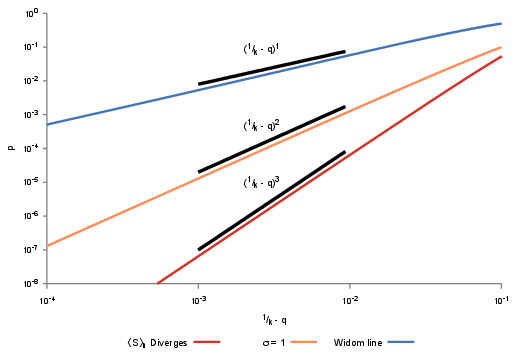}
	\caption{Power-law scaling of the critical and quasi-critical lines near the directed percolation limit. From bottom to top, the critical line, $\sigma =1$, and Widom line in the  limit $p\rightarrow 0$ limit.}
	\label{fig:phasecurve_asymptote}
\end{figure}
In directed percolation, there are several indicators of the critical point. The mean cluster size diverges, the branching ratio is one, and the dynamic susceptibility diverges.  However, with the introduction of spontaneous activation, it is clear that these indicators no longer agree (cf. Fig.~\ref{fig:avroots}\textbf{a} of the main text, or Fig.~\ref{fig:phasecurve_asymptote}). In fact, the branching ratio is no longer a clear signal, because independent streams of activity can merge together and nodes can spontaneously activate. Meanwhile, the dynamic susceptibility no longer diverges, but instead attains a maximum one what is referred to as the Widom line. 

Although the Widom line, unity branching ratio ($\sigma=1$) line, and line of diverging cluster size all agree as $p \rightarrow 0$, they obey different power laws in their approach to that point (Fig.~\ref{fig:phasecurve_asymptote}). In this section, we will derive the different scalings associated with these critical and quasi-critical lines. 

The scaling for the $\sigma = 1$ line is given by $\left(\frac{1}{k} - q \right)^2 \sim p$, which can be seen by solving Eq.~(\ref{eq:pdsimpleeq}) for $q$ and using the closed form for $\Phi$ (Eq.~(\ref{eq:active_fraction_1storder})), which immediately yields $\frac{k^2}{k-2}\left(\frac{1}{k} - q \right)^2 \approx p$ on the $\sigma = 1$ line. 

As for the Widom line, by setting $\pp{\chi_0}{q} = 0$ and applying some simple algebraic manipulation, the Widom line can be found to consist of the $q$ and $p$ satisfying $0 = 1 - k q \overline{\Phi}^2 - 2 \Phi + q \Phi^2$. The first order approximation for $\Phi$ given by Eq.~(\ref{eq:active_fraction_1storder}) is poor in the vicinity of the Widom line (after all, it is in the vicinity of the point of maximum susceptibility in $\Phi$) and so a second order approximation for $\Phi$ (Eq.~(\ref{eq:active_fraction_2ndorder})) is necessary. With this approximation, the Widom line becomes (upon expansion around $p = 0$ and $q = \frac{1}{k}$): $p \approx \frac{k}{2}(\frac{1}{k} - q)$. 

The critical line was previously shown (Eq.~(\ref{eq:phasecurve_scaling})) to obey the scaling $\left(\frac{1}{k} - q \right)^3 \sim p$. These scalings are illustrated in Fig.~\ref{fig:phasecurve_asymptote}.

\section{Phase diagram and scaling collapse on small-world networks \label{sec:appendix_smallworld}}

\begin{figure}[ht]
	\includegraphics[width=\linewidth]{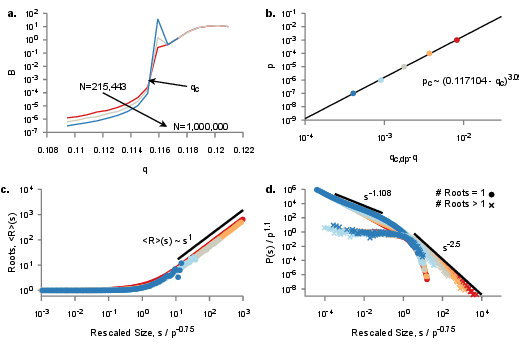}
	\caption{Critical line and avalanche scaling of small-world network with a low rewire probability. The phase diagram and curve collapse along that phase line for $N=10^6$ small-world networks with the re-wire probability $10^{-3}$. \textbf{a} The ratio $B = g \frac{\langle s^2\rangle_c}{\langle s \rangle^2_c}$ for $p = 10^{5}$, identifies the critical point $q_c = 0.115333 \pm 0.000010$.  \textbf{b} The numerically derived critical line is represented with symbols, the black line is the non-linear least-squares fit to the data. \textbf{c} The average number of roots, exhibits a transition that scales with $s_m \sim p^{-0.75}$. \textbf{d} This same $s_m$ effects a curve-collapse in the avalanche distribution, which we see separates the mergeless and merging avalanches. \label{fig:sw_binder_1e-3}}
\end{figure}

By numerically determining the critical point for different $p$ values, we can build a critical line for the small-world networks. In the main-text, we showed that the avalanche distribution $P(s)$ exhibits a scaling collapse when assuming $s_m \sim p^{-2/3}$, for a small-world network with a re-wire probability of $10^{-2}$. This was somewhat surprising, as the small-world network still exhibits a large number of recurrent connections and is very nearly a circulant graph. However, for a lower re-wire probability, $10^{-3}$, recurrent connections play an even larger role, and $p^{-2/3}$ does not provide such a robust scaling. Instead, we find the collapse is best for $p^{-0.75}$ (cf. Fig.~\ref{fig:sw_binder_1e-3}). This can be understood by considering the root-size distribution at the critical point, as is done in  Fig.~\ref{fig:sw_binder_1e-3}\textbf{c}. Clearly, the characteristic merging size $s_m$ scales as $s_m \sim p^{-0.75}$. 

Interestingly, this also allows us to estimate the directed percolation $1/\sigma_{DP}$ exponent for the small-world network. Given a phase-line that scales as $p \sim (q_{c,DP} - q)^a = \delta_{DP}^a$ and a curve collapse effected by $ p^b$, and that we expect the curve collapse to scale as $s_m \sim \delta_{DP}^{1/\sigma_{DP}}$, we have the scaling relation $1/\sigma_{DP} = a b$. The phase diagram for this small-world network (Fig.~\ref{fig:sw_binder_1e-3}\textbf{b}) relates $p \sim (q_{c,DP} - q)^{3.0908}$ for $q_{c,DP} \approx 0.11533$, which implies ${\sigma_{DP}} \approx \frac{1}{3.09  \times 0.75} = 0.43$ for the small-world network with a re-wire probability $10^{-3}$. When the shortcut density is low, the small-world network is approximately a 1-dimensional circulant graph, which suggests we should use the $1+1$-dimensional directed percolation exponents. Our result of $\sigma_{DP} \approx 0.43$ compares reasonably well with the $\sigma_{DP} = 0.391$ reported in the literature \cite{munoz1999avalanche}. Similarly, we should identify the avalanche exponent $\tau_{DP} = 1.108$, which matches well with our numerical results (cf. Fig.~\ref{fig:sw_binder_1e-3}).

\section{Correlation length \label{sec:appendix_Correlation_length}}
In the main text, we introduce the pair connectedness function $g(d,t)$. By consider the (typically exponential) decay of this pair-connectedness function, we can define correlation lengths $\xi$ of the form $g(d) \sim \exp[-d/\xi]$. We will show analytically that the perpendicular correlation length $\xi_\perp$ corresponding to the decay of $g(d,0)$  diverges on the critical line. 

\subsection{Divergence of perpendicular correlation length}
We can derive the divergence of the perpendicular correlation length $\xi_\perp$ by computing $g(2d,0)$. We consider $2d$, because when a daughter branch is followed, $t$ advances by one, while a parent branch decreases $t$ by one. However, $\Delta t = 0$, so the number of parent and daughter branches must both be equal.

\begin{figure*}
	\includegraphics[width=\linewidth]{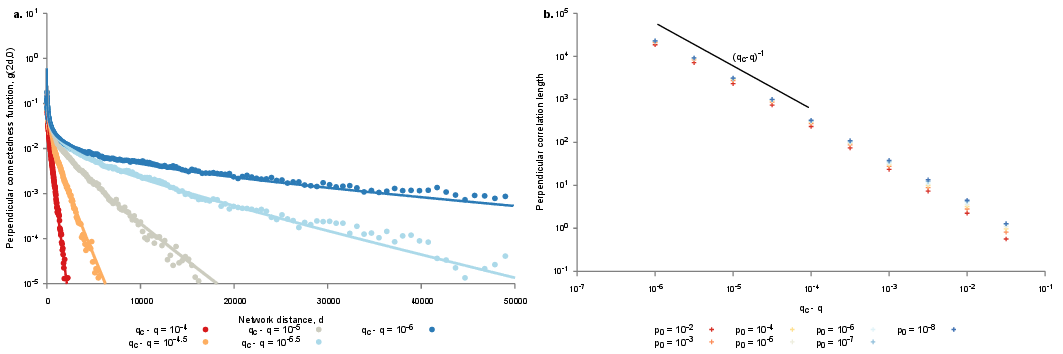}
`    \caption{Perpendicular correlation length near criticality on infinite $k$-regular networks. \textbf{a} The simultaneous (perpendicular) path connectedness function has an exponentially decaying tail, with the exact form predicted analytically. Solid lines are analytical results while symbols are numerical simulations on infinite 10-regular networks (averaged over 20,000,000 clusters), simulated at $p_0 = 10^{-3}$.  \textbf{b} The isotropic correlation length diverges with a power-law of $(q_c-q)^{-1}$. \label{fig:correlation_perp}}
\end{figure*}

That being said, not all routes with $d$ daughters and $d$ parent connections are  are equally likely. For instance, whenever a  parental connection follows a daughter connection, the route requires that two initially independent avalanches merge at that point. It turns out that the most convenient way to compute $g(2d,0)$ is to sum over collections of routes that have a fixed number of merges. The number of routes with $m$ merges is given by \begin{equation}\left[\binom{d}{m} + \frac{1}{k-1}\binom{d-1}{m-1}\right]^2k^{2d-1}\left(\frac{k-1}{k}\right)^{2m} \,.\end{equation} Given a sequence of parent / daughter network hops, the combinatorial factor $\binom{d}{m}$ counts the number of ways that the parent / daughter network hops could be rearranged without altering the number of merges. A network hop that follows a network hop of the same kind (i.e. a parent hop following a parent hop, or a daughter hop following a daughter), contributes $k$ possible paths. Every time a change in direction occurs (i.e. a daughter followed by parent or parent by daughter), only $k-1$ links are available, because one was taken to arrive at the node in question. The factor of $k^{2d-1}\left(\frac{k-1}{k}\right)^{2m}$ captures the number of possible paths, given a sequence of daughter / parent network hops. A correction of $\frac{1}{k-1}\binom{d-1}{m-1}$ is required, to account for those paths with one or two fewer change in direction (a boundary condition effect).  A daughter or parental connection occurs with weight $P_d$ or $P_p = P_d$, except when a parental connection follows a daughter connection, when it instead contributes $P_{p1}$. So, since the number of merges could range from $0$ to $d$ we can write:
\begin{widetext}
\begin{align*}
g(2d,0) &= \sum_{m=0}^d  \Bigg(\left[(k-1)\binom{d}{m} + \binom{d-1}{m-1}\right]^2 \times \left(\frac{k-1}{k}\right)^{2m-1} \frac{(k P_d)^{2d}}{k^2}\left(\frac{P_{p1}}{P_d}\right)^m \Bigg).\\
\intertext{
This expression is compared to simulations on the infinite lattice in Fig.~\ref{fig:correlation_perp}, and has the closed form expansion:}
g(2d,0)\frac{k \sigma -(k-1)^2 P_{p1}}{\sigma^d}&= -\bigg(2(\sigma - (k-1)P_{p1})\hypergeometricF(1-d,-d;1;\beta) +((k+1)\sigma + kP_{p1}) \hypergeometricF(1-d,1-d;1;\beta)  \bigg),
\end{align*}
\end{widetext}
where $\hypergeometricF$ denotes the Gauss hypergeometric function, and $\sigma = k P_d$ and $\beta = \frac{(k-1)^2 P_{p1}}{k^2 P_d}$ as before. Since in the limit of large $d$, $\frac{\hypergeometricF(1-d,-d;1;\beta)}{\hypergeometricF(1-(d-1),-(d-1);1;\beta)} = \frac{\hypergeometricF(1-d,1-d;1;\beta)}{\hypergeometricF(1-(d-1),1-(d-1);1;\beta)}$, we need only treat one of them asymptotically. By way of Kummer's 24 solutions \cite{luke1969special}, we have that $\hypergeometricF(1-d,-d;1;\beta) = (1-\beta)^d\hypergeometricF(d,-d;1;\beta / (1-\beta))$. Using an identity owing to Wilson \cite{wilson1918asymptotic} we have $\hypergeometricF(1-d,-d;1;\beta) \propto \frac{(1+\sqrt{\beta})^d}{\sqrt{d}}$, and so in the limit of large $d$,
\begin{equation}
\frac{g(2(d+1),0)}{g(2d,0)} \approx \left(\sigma (1+\sqrt{\beta})\right)^2,
\end{equation}
which yields $\xi_\perp = \frac{-1}{2 \log(\sigma (1+\sqrt{\beta}))} \sim (q_c-q)^{-1}$, yielding the scaling exponent $\nu_\perp = 1$. Finally, it's clear that the divergence in correlation length occurs precisely when $\sigma^2 \left(1+\sqrt{\beta}\right)^2 = 1$. Simple algebra shows that this is equivalent to $\sigma^2 \beta = (1-\sigma^2)$, which is the critical line derived by considering the divergence of the average cluster size (see Eq.~\ref{eq:critical_line_beta}). 

\begin{figure}[htbp]
	\includegraphics[width=\linewidth]{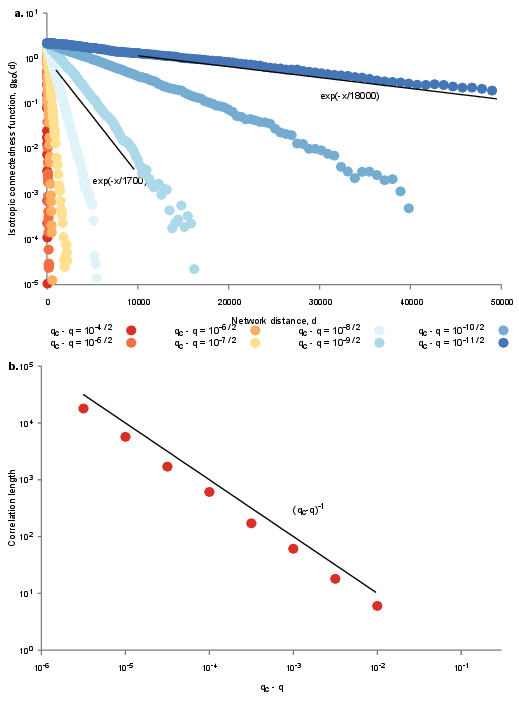}
	\caption{Isotropic correlation length near criticality foron infinite $k$-regular networks. \textbf{a} The istropic path connectedness function $g_{iso}(d)$ for all paths of length $d$ decays exponentially. This is the result of simulations of infinite 10-regular networks, simulated at $p = 10^{-3}$.  \textbf{b} The corresponding isotropic correlation length $\xi$ diverges with a power-law of $(q_c-q)^{-1}$. \label{fig:isotropic_divergences} }
\end{figure}

\subsection{Alternative correlation lengths}
An alternative parallel correlation length $\xi_\parallel$ for random graphs is the length characterizing the decay of direct descendants of an active site. This can be measured with $g(t,t) = \sigma^t = \exp[-t / \xi_d]$ where $\xi_d = -1/\ln(\sigma)$ denotes the descendant correlation length. This correlation length is given strictly by the branching ratio, $\sigma$, and diverges as $\xi_d \sim |q(\sigma = 1, p) - q|^{-1}$. However, the $\sigma = 1$ line on which  $\xi_d$ diverges is well into the super-critical regime, save for the singular point $p = 0$ and $q = \frac{1}{k}$.

We've considered a correlation length $\xi_\perp$ that corresponds to the directed percolation perpendicular correlation length. $\xi_\perp$ is anisotropic, which may appear to make it unrelated to the isotropic correlation length of undirected percolation. However, the presence of a diverging anistropic correlation length implies that any isotropic correlation length will also diverge. To illustrate this, consider the correlation length, $\xi$ defined by the mean number of sites active after $d$ (isotropic) network hops away from an active node, i.e.  $g_{iso}(d) = \sum_{t=-d}^d g(d,t) \sim \exp[-d/\xi]$. Since this sum contains $g(d,0) \sim \exp[-d / (2\xi_\perp)]$, which tends to a constant as $q\rightarrow q_c$, we know therefore that $\xi$ also diverges in the same limit, as can be seen in Fig.~\ref{fig:isotropic_divergences}.

\section{Static model \label{sec:appendix_SIR}}
\begin{figure}[htbp]
    \centering
    \includegraphics[]{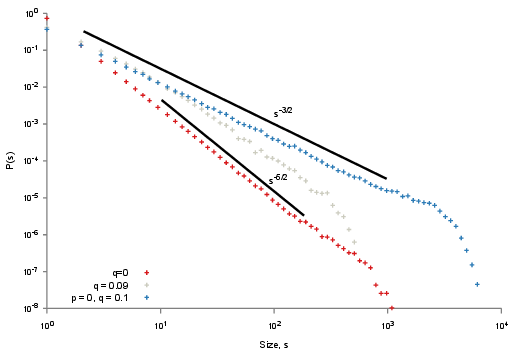}
    \caption{Cluster size distributions on undirected 10-regular networks with $N=2^{16}$. $p$ fraction of nodes are infected and disease spreads with the probability $q$ through the links, which corresponds to SIR model. Only when $p=0$, do we consider a single patient zero. Data points are the result of  1000 realizations, except $p=0$, $q=0.1$ which has 100,000 realizations.
    \label{fig:sw_static_model}}
\end{figure}

Many diseases exhibit immunity after spreading. In this paper, we have predominately considered a model with re-excitable nodes, corresponding to the SIS model. It is, however, also of interest to consider diseases that might have multiple initiation points, but which cannot reinfect an individual after they have contracted the disease. This is a modification of the canonical SIR model. To consider such diseases, we initially infect some fraction $p$ of the network. Each connection between nodes transmits the infection with probability $q$. The analogy to our initial model is not exact because we have disposed of the temporal aspect of the model, i.e. \textit{de novo} infections are not repeatedly introduced to the system. Nonetheless, there are two modes by which the disease grows -- a period of spreading followed by the merging of large clusters. Unsurprisingly, this model also shows two power-laws in the cluster size, with an exponent of $-2.5$ for $q \rightarrow 0$ and an exponent of $-1.5$ for $p \rightarrow 0$ (cf. Fig.~(\ref{fig:sw_static_model})) on directed $k$-regular networks in the percolation and directed percolation limits respectively. 

In contrast to our results with the SIS model, we anticipate that in SIR this transition is more sensitive to the details of the network. Immunization plays a role when activity traverses a loop in the network. Percolation on directed $k$-regular networks is mean-field, and so loops play little role. However, for networks where loops are prevalent (such as small-world networks, or undirected networks) immunization may have a stronger effect. It is known, for instance, that the SIR model falls into the dynamical percolation universality class and shows an undirected percolation phase transition on undirected networks  \cite{tome2010critical,newman2002spread}. The directed percolation exponents are only demonstrated with SIR on directed networks  \cite{schwartz2002percolation}. It would be interesting to see which aspects of the spreading-merging transition remain, if the SIR model were to be simulated on undirected networks.

\section{Additional figures}
\begin{figure*}[htbp]
	\includegraphics[width=\linewidth]{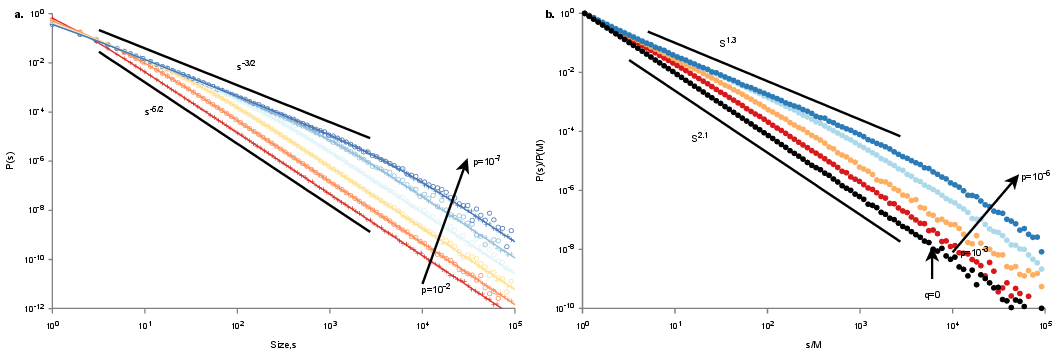}
	\caption{Unscaled critical avalanche distributions. \textbf{a} Avalanche distributions for 10-regular networks as in Fig.~\ref{fig:avdists}\textbf{e}. Solid lines are analytical $P(s)$ determined from the generating function, while the crosses and circles are simulations on infinite and finite ($N=10^7$) networks. \textbf{b} Avalanche distributions for HMN networks as in Fig.~\ref{fig:avdists}\textbf{f} of the main text. \label{fig:unscaled_avalanches}}
\end{figure*}

\begin{figure}[htbp]
	\includegraphics[width=\linewidth]{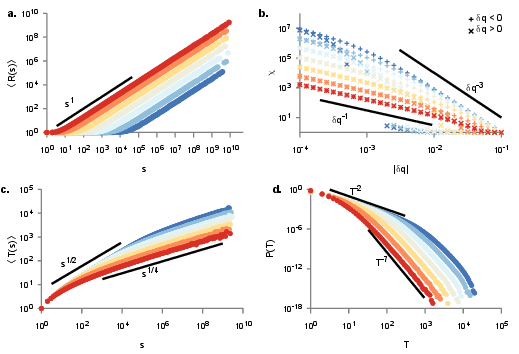}
	\caption{Exponent transitions for critical avalanches, without the curve collapse presented in the main text. Exponent transitions for $10$-regular networks, without the rescaling presented in Fig.~\ref{fig:avroots} of the main text. Data are for $p = 10^{-2}$ to $p = 10^{-8}$. \textbf{a} The average number of roots for avalanches of a given size, as simulated on infinite networks. \textbf{b} The analytically determined susceptibility $\chi$. \textbf{c} The avalanche duration-size relation.  \textbf{d} The avalanche duration distribution.  \label{fig:unscaled_rollovers}}
\end{figure}

In this section we include several figures to supplement the main text. Fig.~\ref{fig:unscaled_avalanches} contains un-scaled avalanche distributions for the hierarchical modular networks and 10-regular networks corresponding to Fig.~\ref{fig:avdists}. Similarly, Fig.~\ref{fig:unscaled_rollovers} consists of the exponent transitions for 10-regular graphs without the exponent transitions presented in Fig.~\ref{fig:avroots}. 

\bibliography{biblio.bib,exportTimeScales.bib,criticalBrain.bib}

\vskip 0.5cm

\noindent

\end{document}